\documentclass[
    onecolumn,
    preprintnumbers,
    amsmath,
    amssymb,
    prd,
    showpacs
]{revtex4}
\usepackage{graphicx}%
\usepackage{amsmath}
\usepackage[normalem,normalbf]{ulem}
\usepackage{amssymb}
\usepackage{bm}
\usepackage{enumerate}

\begin{document}
\title{Dynamics of phase transition
   in (3+1) dimensional scalar $\phi^4$ theory}
\author{Hyeong-Chan Kim}
\email{hckim@phya.yonsei.ac.kr}
\author{Jae Hyung Yee}%
\email{jhyee@phya.yonsei.ac.kr}
\affiliation{Institute of Physics and Applied Physics, Yonsei
University, Seoul, Republic of Korea.
}%
\date{\today}%
\bigskip
\begin{abstract}
\bigskip

We use the variational approximation with double Gaussian type
trial wave-functional approximation, in which we use the square
root of the dispersion of the zero-mode wave-function as an order
parameter, to study the out of equilibrium quantum dynamics of
time-dependent second order phase transitions in (3+1) dimensions.
We study the time evolution of symmetric states of scalar $\lambda
\phi^4$ theory in several situations by properly treating the
effect of the $\lambda\phi^4$ interaction. We also calculate the
effective action and the effective potential of the theory with
the precarious renormalization.  We show that the presence of a
quenching of the mass-squared leads to second order phase
transition nontrivially since the vacuum structure changes by
absorbing the energy required for quenching, even though there is
no symmetry breaking in the effective potential of the theory
without quenching process. We also calculate the equal time
correlation function, and then evaluate the correlation length as
a function of the mass-squared. The time dependence of the
correlation length varies depending on how the mass-squared
changes in time. For constant mass-squared it gives the classical
Cahn-Allen relation, and it leads to different relations for other
time-dependence of the mass-squared. We also show that there
exists a propagating spatial correlation after termination of the
phase transition process in addition to the correlation
corresponding to the formation and growth of domains.

\end{abstract}
\pacs{11.15.Tk, 05.70.Ln}
\keywords{Symmetry breaking, beyond Gaussian approximation}
\maketitle
\newpage

\section{Introduction}

In recent years, there has been strong interest in the dynamics of
the quantum fields out of equilibrium appearing in the evolution
of the  early universe or in the process of phase
transition~\cite{guth,vilenkin}. The formation of topological
defects, the dynamics of inflation of the universe, and the
inflationary reheating have been studied on the basis of scalar
field model. The domain formation and growth~\cite{boyanovsky}
were studied in the out of equilibrium second order phase
transition using the Hartree-Fock approximation. During a second
order phase transition, the time scale of relaxation of the scalar
field lags behind the time scale of change of the effective
potential. Consequently, the field evolves out of equilibrium as
it tries to relax to a new vacuum, giving rise to a nonvanishing
vacuum expectation value. Such nonequilibrium effects play a
crucial role in topological defect formation in condensed matter
systems as well as in the early
universe~\cite{gill,ibaceta,stephens}. Kibble first showed how the
correlation length is crucial in determining the initial density
of topological defects~\cite{kibble}. These ideas were elaborated
by Zurek who proposed that it may be possible to quantitatively
test the Kibble mechanism of defect formation in condensed matter
systems such as superfluid He$^4$~\cite{zurek}. He argued that,
because of the phenomenon of critical slowing down near the phase
transition point, the correlation length relevant for determining
the initial density of the defects is not the equilibrium
correlation length at the Ginzburg temperature but that at the
time when the field dynamics essentially freezes out. He also
found a power law behavior in the dependence of the correlation
length on the quench rate. There have been various attempts to
experimentally test the Kibble-Zurek prediction of initial defect
density~\cite{laguna}. A comprehensive review on these issues is
given in Ref.~\cite{rivers}.

One crucial lesson provided by these studies is the importance of
using time-dependent techniques to study processes of the system
with explicit time-dependence. It has been repeatedly shown that
classical and one-loop effective potentials are poorly defined and
of little use in such dynamical systems. Another common theme is
the importance of non-linear corrections to the linear dynamics.
In these directions, the dynamics of spinodal decomposition in
inflationary cosmology using the closed time path formalism of
quantum field theory out of equilibrium combined with the
non-perturbative Hartree approximation was analyzed in
Ref.~\cite{cormier}. A feature common to all such phase
transitions is the evolution far away from equilibrium and the
exponential growth of the soft (long wavelength) modes, which
necessarily lead to domain growth. Finite temperature field theory
based on equilibrium or quasi-equilibrium methods does not
describe all the processes of nonequilibrium evolution even though
the imaginary part of the complex effective action yields the
decay rate~\cite{weinberg}. To treat such nonequilibrium quantum
evolution properly, Schwinger and Keldysh first introduced the
closed-time formalism~\cite{schwinger}. The closed-time formalism
and the $1/N$ expansion method have been applied to nonequilibrium
$\lambda\phi^4$ theory to explain the phenomenon of domain
growth~\cite{cooper}. Recently, the Liouville-von Neumann approach
has been developed that unifies both the functional
Schr\"{o}dinger equation for quantum evolution and the quantum
Liouville equation for quantum statistical mechanics~\cite{spkim}.
The renormalization of field theory in quenched second order phase
transitions was discussed in Ref.~\cite{spkim2} in this context.

An obstacle in discussing the phase transition of the scalar field
in (3+1) dimensions is that there is no symmetry breaking in the
Gaussian effective potential of the theory with the precarious
renormalization~\cite{stevenson}, which is the only relevant
renormalization scheme which does not require non-trivial
assumptions such as the existence of a large momentum cutoff,
large $N$ limit of $O(N)$ scalar field, or the autonomous
renormalization~\cite{stevenson2,jhyee}. On the other hand, there
have been various studies on the 2nd order phase transitions of
the scalar $\phi^4$ theory in (3+1)dimensions in relation to the
$O(N)$ symmetric theory~\cite{cooper}, the non-equilibrium field
theories~\cite{spkim2,cormier}, and the theory with high momentum
cutoff~\cite{consoli}. Most of these studies consider the region
in which the phase transition begins, where the dynamical behavior
of the unstable modes are the main interest. It is an interesting
question to which vacuum the system will settle down after the
transition since there is no vacuum with non-zero order parameter
in the static theory described by the effective potential. In this
paper we answer this question by showing that the vacuum structure
of the scalar field theory changes when the second order phase
transition occurs.

The variational approach for the scalar field theory using the
Gaussian effective potential was well studied in
Refs.~\cite{stevenson,cea,barns}, and references therein. The
renormalizability and the initial value problem for the Gaussian
approximation of time-dependent scalar systems are checked in
~\cite{cooper3,pi1}. Many authors have studied the symmetry
breaking phase structures in the large $N$
approximation~\cite{baacke}. Non-equilibrium dynamics of symmetry
breaking~\cite{cooper2} and the second order phase
transition~\cite{spkim} have also been studied in the Gaussian
frame work. The Hartree-Fock method has been popular and useful in
studying the nonequilibrium evolution of the field~\cite{cooper}.
Even though the renormalization is well understood even for
nonequilibrium fields in Refs.~\cite{cooper} a systematic and
explicit renormalization scheme of the effective action has not
been properly addressed for the systems undergoing phase
transitions at least in (3+1) dimensions. In this paper, we
propose such a renormalization scheme of the effective action for
systems describing phase transitions.

As field theory models for the second order phase transition, some
exactly solvable models of the free scalar field were studied in
Ref.~\cite{spkim}, in which it was proposed that the exponential
growth of the spinodal instability may ends at the spinodal line
because of the back-reaction of the scalar field. The spinodal
instability leads to the domain formation and
growth~\cite{boyanovsky}, which is determined by the equal time
correlation functions. The formation and growth of the domains
were studied up to the point where the phase transition process
ends. It is generally believed that the domain becomes static once
it is formed without other dynamics involved. In this paper, by
studying the spatial correlation functions determined by the
approximation method proposed in Ref.~\cite{kim}, we show that
there exists another propagating correlations after the transition
process is over in addition to the static domain.

Recently, the present authors have developed a new quartic
exponential type variational approximation~\cite{kim1} which is
suitable for double well type potentials in the quantum mechanical
context. This approach was applied to the zero mode of the scalar
field~\cite{kim}, where we have renormalized the equations of
motion which describes the symmetry breaking phenomena starting
from the symmetric states of the scalar $\lambda\phi^4$ theory in
three and four dimensions. In the present paper, we use some of
these results summarized in Sec. II.

The article is organized as follows. In Sec. II, we briefly
present the double Gaussian type wave-functional approximation to
the $\lambda\phi^4$ theory. Then, we introduce the Klein-Gordon
like mode solutions, with which we rewrite the Hamiltonian of the
system so that it includes the temperature of the initial
equilibrium state. In Sec. III, we calculate the effective action
and potential of the interacting scalar field with fixed mass in
which the tachyonic mode described by negative mass-squared is
included. It is shown that the typical effective potential
develops a new local minima at the non-vanishing expectation value
of the field at which point the mass vanishes even though the
vacuum still presents at the symmetric point of order parameter.
Sec. IV is devoted to the analysis of an exact mode solution of
the field equation in which the mass-squared varies from positive
to negative values. We obtain several asymptotic behaviors of the
solution such as the instantaneous quenching, $t\rightarrow
\infty$ limit, and ultra-violet (UV) limit. These are used in Sec.
V to obtain the WKB approximation for the general field equation.
In Sec. V, we study the self-interacting scalar field system with
instantaneous quenching at $t=0$. After the quenching we allow the
mass-squared freely evolve so that it increases due to the
dynamical back reaction of $\lambda \phi^4$ interaction. We
calculate the UV finite renormalized Hamiltonian and potential
written in terms of a shifted order parameter, $\bar{q}$, and
mass-squared, $m^2$, with an additional relation between $\bar{q}$
and $m^2$. In Sec. VI, we introduce a large instability
approximation, with which we analytically study the effective
Hamiltonian and potential. It is shown that the vacuum structure
of the theory is changed due to the presence of instantaneous
quench at $t=0$ so that there occurs second order phase transition
in the new effective potential. In Sec. VII, we calculate the
evolution of the equal time correlation functions and present a
simple formula which relates the correlation length and the
time-dependent mass-squared. We also show that there exists a
propagating correlation in addition to the usual correlation which
describes formation of domains. Summary of our results and some
discussions are given in Sec. VIII, and three appendices are added
at the end of the article.

\section{Mode separation of the self-interacting scalar field}
In this section, we summarize the double Gaussian type
approximation of the self-interacting scalar field
theory~\cite{kim}, in which the back reaction of the field by the
$\lambda\phi^4$ interaction is considered in a natural fashion.
Then, we briefly describe how the time evolution of the initial
equilibrium state with inverse temperature $\beta$ can be analyzed
in the present framework.

The Lagrangian of self interacting scalar theory in $3+1$
dimensions is
\begin{eqnarray} \label{L}
L(t)= \int d^3 {\bf x} \left[ \frac{1}{2} \partial^\mu \phi(x^\nu)
\partial_\mu \phi(x^\nu) - \frac{1}{2}\mu^2(t) \phi^2(x^\nu)
 -\frac{\lambda}{4!} \phi^4(x^\nu)  \right] ,
\end{eqnarray}
where we explicitly included the volume integral so that we can
write the volume factor in the zero mode part of the Lagrangian.
In Ref.~\cite{kim} we have shown that the quantum mechanical
excitation of the zero mode may play a non-trivial role in the
symmetry breaking of the system, where we separated the zero mode
from the other modes by the following form,
\begin{eqnarray} \label{phi0}
\phi(t) = \frac{\int d^3 {\bf x} \phi({\bf x},t)}{\int d^3 {\bf
x}}, ~~~ \psi({\bf x},t) = \phi({\bf x},t) -
    \phi(t) .
\end{eqnarray}
We use the unit $\hbar=1$ in this paper. Following the double
Gaussian type approximation to the zero mode developed in
Ref.~\cite{kim1}, we take the trial wave-functional
\begin{eqnarray} \label{wf1}
\Psi[\phi,\psi]=N \exp \left\{-\frac{1}{2} \left[\frac{1}{2g^2(t)}
+ i \pi (t) \right]\phi^4+ \left[\frac{x(t)}{g(t)}+
ip(t)\right]\phi^2-\int_{\bf xy} \psi({\bf x})
\left[\frac{G^{-1}({\bf x,y};t)}{4}-i \Pi({\bf x,y};t)
\right]\psi({\bf y})\right\},
\end{eqnarray}
where $\int_{\bf x}= \int d^3 {\bf x}$. Then the
Lagrangian~(\ref{L}) leads to the effective action
\begin{eqnarray} \label{Gamma:3}
S[q,y;G,\Pi] &=& \int dt d^3{\bf
x}\left\{\left[\frac{q^2D^2(y)\dot{y}^2}{2}
     + \frac{\dot{q}^2}{2} - V(q,y)\right] +
  \int_{\bf k}\left[\Pi({\bf k},t) \dot{G}({\bf k},t)
      -2  \Pi^2({\bf k},t) G({\bf k},t)
   \right. \right. \\
&-& \left. \left.
  \frac{1}{8}G^{-1}({\bf k},t) - \frac{1}{2}
   \left({\bf k}^2
 + \mu^2+\frac{\lambda}{2} q^2  \right) G({\bf k},t)\right] -
  \frac{\lambda}{8}\left[
   \int_{\bf k} G({\bf k},t) \right]^2 \right\} , \nonumber
\end{eqnarray}
where $D(y)$ is a smooth function of $y$ defined in
Ref.~\cite{kim},
\begin{eqnarray} \label{dga}
q(t)&=& \sqrt{\langle \phi^2(t) \rangle}, ~~ y(t)= \frac{\langle
\phi^4(t) \rangle}{\langle \phi^2(t) \rangle^2}
\end{eqnarray}
describe the dispersion and shape of the whole double Gaussian
type wave-functional of the zero mode, respectively, and
$\displaystyle \langle \psi({\bf x},t)\psi({\bf y},t) \rangle=\int
\frac{d^3{\bf k}}{(2\pi)^3} e^{i{\bf k}\cdot ({\bf x-y})}G({\bf
k},t)$ denotes the width of the Gaussian trial wave-functional for
the non-zero modes. The zero-mode potential is
\begin{eqnarray} \label{Veff}
V(q,y) =  \frac{1}{2}\mu^2 q^2 + \frac{\lambda}{4!}
     y q^4,  ~~y\geq 1, ~~q \geq 0 .
\end{eqnarray}
Note that the potential $V$ is finite even in the limit of zero
dispersion, $q^2 \rightarrow 0$ or $y \rightarrow 1$, which is the
distinct feature of the theories of infinite volume. The
time-dependent variational equations are given by
\begin{eqnarray} \label{eqs}
&& \frac{d}{dt}(q^2 D\dot{y})+ \frac{\lambda}{4! D}q^4=0 , \\
   \label{ddq:0}
&&\ddot{q}+[m^2(t)-\dot{\eta}^2(t)]q+ \frac{\lambda}{6}(y-3)q^3=0, \\
&& \ddot{G}({\bf k},t)= \frac{1}{2}G^{-1}({\bf k},t)+ \frac{1}{2}
G^{-1}({\bf k},t)\dot{G}^2({\bf k},t)-2
  [{\bf k}^2+ m^2(t)] G({\bf k},t) , \label{gap:1}
\end{eqnarray}
where
 \begin{eqnarray} \label{m:qG}
 m^2(t) &=& \mu^2(t)+
\frac{\lambda}{2} q^2(t)+ \frac{\lambda}{2} \int_{\bf k} G({\bf k},t) .
\end{eqnarray}
A distinctive feature of this equation is that the square root of
the dispersion of the zero mode, $q$, satisfies the deformed
classical equation of motion~(\ref{ddq:0}), and can play the role
of order parameter.

The potential divergences in the integral $\int_{\bf k}G({\bf
k},t)$ of the equations of motion are absorbed into the bare mass
$\mu^2$ and the bare coupling $\lambda$ leading to the
renormalized ones. We have shown in the previous paper~\cite{kim}
that the equations of motion are renormalizable. But the proof is
not complete in the following sense: 1) the theory considered in
the previous paper does not include the unstable modes where
negative mass-squared appears and spinodal instability grows. 2)
the proof does not include the renormalizability of the effective
action even though it shows the renormalizability of equations of
motion. In computing the effective action, we have extra divergent
integrals of the form $\int_{\bf k}G^{-1}_{\bf k}$ and $\int_{\bf
k} \omega_{\bf k}^2 G({\bf k},t)$ which may cause extra
complications. For simplicity we set $y=1$ in this paper so that
it is non-dynamical. The dynamics of $y$ may give non-trivial
contributions to the effective action, however, its inclusion is
not difficult because dynamical equation~(\ref{eqs}) for $y$
simply relates $\dot{y}$ with $1/q^2$ in the precarious
renormalization scheme.

To consider the spinodal instability, we rewrite the equation for
$G_{\bf k}$ by introducing the mode solution $\phi_{\bf k}(t)$ of
the Klein-Gordon equation,
\begin{eqnarray} \label{ddphi:0} \ddot{\phi}_{\bf k}(t) +
\omega^2_{\bf k}(t) \phi_{\bf k}(t)=0 ,
\end{eqnarray}
where $\phi_{\bf k}= R_{\bf k}e^{i\theta_{\bf k}(t)}$. With the
identifications,
\begin{eqnarray} \label{psi:k}
R_{\bf k}^2 \dot{\theta}_{\bf k} &=& \frac{1}{2},
\nonumber\\
\omega_{\bf
   k}^2(t)&=& {\bf k}^2 + \mu^2(t) + \frac{\lambda}{2} q^2(t) +
   \frac{\lambda}{2} \int_{\bf k} G_{\bf k} ,
   \label{omega:k}
\end{eqnarray}
we obtain
\begin{eqnarray} \label{G:psi}
G({\bf k},t) &=& \phi_{\bf k}^*(t) \phi_{\bf k}(t).
\end{eqnarray}

For time-independent system with fixed mass $m(t)=m_i>0$, the
solution for $\phi_{\bf k}(t)$ is given by
\begin{eqnarray} \label{phi:i}
\phi_{i, \bf k}(t)= \frac{1}{\sqrt{2 \omega_{i, \bf k}}}e^{-i
\omega_{i , \bf k}t} ,
\end{eqnarray}
where $\omega_{i,{\bf k}}= \sqrt{{\bf k}^2 + m_i^2}$ is the
initial frequency of mode $\phi_{\bf k}$. The direct link of the
mode solution and the Gaussian approximation was discussed in
Ref.~\cite{kim3} and this Klein-Gordon type mode solution was
discussed in Ref.~\cite{spkim} in the context of Liouville-von
Neumann approach to the scalar field theory. As in the case of the
reference~\cite{spkim}, we can define the thermal expectation
value at $t=0$, which gives the Hamiltonian density
\begin{eqnarray} \label{H}
H(q,p;\dot{G},G)
 &=&\frac{\dot{q}^2}{2}+ \frac{1}{2}\mu^2q^2(t)
 + \frac{\lambda}{4!}q^4+ \int_{\bf k}H_{\bf k}(t)
   \coth \left(\frac{\beta \omega_{i,\bf k}}{2} \right) -
  \frac{\lambda}{8}
   \left[\int_{\bf k} \phi_{\bf k}^*(t) \phi_{\bf k}(t)
   \coth \left(\frac{\beta \omega_{i,\bf k}}{2}
      \right)\right]^2,
\end{eqnarray}
where $\beta$ is the inverse temperature at $t=0$ and
\begin{eqnarray} \label{Hk}
H_{\bf k}(t)&=&\frac{1}{2}\left[ \dot{\phi}_{\bf k}^*(t)
\dot{\phi}_{\bf k}(t) +
   \omega_{\bf k}^2(t) \phi_{\bf k}^*(t)\phi_{\bf k}(t)\right] , \\
\omega_{\bf k}^2(t)&=& {\bf k}^2 + \mu^2(t) + \frac{\lambda}{2}
q^2(t) +
   \frac{\lambda}{2} \int_{\bf k} \phi_{\bf k}^*(t) \phi_{\bf k}(t)
   \coth\left(\frac{\beta \omega_{i, {\bf k}}}{2}\right)
   \label{omega:kT}  .
\end{eqnarray}
The generalization to thermal state comes from the equivalence of
the Liouville-von Neumann equation and the Gaussian
wave-functional approach, which was well described in
Ref.~\cite{spkim}. We do not digress about this point in this
article. This Hamiltonian can be compared with Eq.~(7.18) in
Ref.~\cite{spkim}. Note that the frequency in the presence of
temperature includes the temperature dependence in the momentum
integral.

\section{Effective action for the case with
constant mass-squared including the Tachyonic modes}
Since solving Eqs.~(\ref{eqs}), (\ref{ddq:0}), and (\ref{ddphi:0})
is not simple, we start from the simplest case, the one with the
mass-squared $m^2(t)=m^2$ being fixed irrespective of its sign.
Even though, in realistic system, the mass-squared may change by
the $\lambda \phi^4$ interactions, we assume that the mass-squared
is kept fixed in this section by an external mechanism which we do
not specify. This procedure may need an external work, which may
alter the energy of the system. For simplicity of the calculation,
we calculate only the zero temperature case. In the subsequent
sections, we calculate the case of time-varying mass-squared
ignoring the self interaction, and then we synthesize the two
cases to calculate the effective action for a self-interacting
scalar field theory with time-dependent symmetry breaking.

We first define an integral notation, $\int_{+,\bf k} f_{\bf
k}(t)$, by
\begin{eqnarray} \label{int:def}
\int_{+,\bf k} f_{\bf k}(t)
= \left\{
  \begin{tabular}{ll}
    $\displaystyle \int_{|{\bf
     k}|=\bar{m}}^{\infty} \frac{d^3 {\bf k}}{(2\pi)^3}f_{\bf
      k}(t)$,& ~~$
      -m^2(t)=\bar{m}^2(t) >0$, \vspace{.1cm}\\
   $\displaystyle \int_{{\bf k}=0}^{\infty} \frac{d^3 {\bf
    k}}{(2\pi)^3}f_{\bf k}(t)$, & ~~ $m^2(t) \geq 0$. \\
   \end{tabular} \right.
\end{eqnarray}
We always use the notation: $\bar{m}^2= |m^2|$ if $m^2$ is
negative. During the calculation of the effective action we
frequently encounter integrals of the form:
\begin{eqnarray} \label{I:def}
I_N(m^2)=\frac{1}{2}\int_{+,\bf k} ({\bf k}^2 + m^2)^{N-1/2},
\end{eqnarray}
where we keep the square in the argument since the mass-squared
can be negative. In the case of positive definite mass-squared,
$m^2>0$, this integrals are well studied by
Stevenson~\cite{stevenson} in which the reduction formula of the
divergences are given by
\begin{eqnarray} \label{stevenson:fmr}
I_1(m^2)-I_1(m_R^2)&=&\frac{1}{2}(m^2-m_R^2)I_0(m_R^2)-
\frac{1}{8}(m^2-m_R^2)^2I_{-1}(m_R^2)+ \frac{m_R^4}{32\pi^2}L_3(x), \\
I_0(m^2)-I_0(m^2_R)&=&-\frac{1}{2}(m^2-m_R^2)I_{-1}(m_R^2)+
\frac{m_R^2}{16\pi^2}L_2(x), \nonumber \\
I_{-1}(m^2)-I_{-1}(m_R^2)&=&-\frac{1}{8\pi^2}\ln x, \nonumber
\end{eqnarray}
where $\displaystyle x=\frac{m^2}{m_R^2}$, and
\begin{eqnarray} \label{Ls}
L_2(x)= x \ln x- x+1,~~L_3(x)= \frac{1}{4}[2x^2 \ln
x-2(x-1)-3(x-1)^2] .
\end{eqnarray}
Beside the above formula we need the following two formula which
relates the integrals with positive and negative mass-squared,
\begin{eqnarray} \label{formula}
I_1(\bar{m}^2)-I_1(-\bar{m}^2)&=& \bar{m}^2\left[I_0(\bar{m}^2)+
\frac{1}{2}\bar{m}^2
I_{-1}(\bar{m}^2)\right]+\frac{\bar{m}^4}{16\pi^2}, \\
I_0(\bar{m}^2)-I_0(-\bar{m}^2) &=& -\bar{m}^2I_{-1}(\bar{m}^2)-
\frac{\bar{m}^2}{8\pi^2} .\nonumber
\end{eqnarray}

Armed with these formula~(\ref{stevenson:fmr}) and
(\ref{formula}), we are prepared to calculate the effective
action. The Klein-Gordon equation~(\ref{ddphi:0}) can easily be
solved to give
\begin{eqnarray} \label{phi:T}
\phi_{\bf k}(t) &=& \frac{1}{\sqrt{2\tilde{\omega}_{\bf k}}}
\left[c_{1\bf k} e^{\bar{\omega}_{\bf k} t}+c_{2 \bf k}
e^{-\bar{\omega}_{\bf k} t}\right],
~~ 0\leq {\bf k}^2 < -m^2, \\
\phi_{\bf k}(t) &=& \frac{1}{\sqrt{2\omega_{\bf k}}} e^{-i
\omega_{\bf k} t}, ~~ {\bf k}^2 \geq -m^2>0 , \nonumber
\end{eqnarray}
where $\tilde{\omega}^2_{\bf k}=-m^2-{\bf k}^2 $ for ${\bf
k}^2<-m^2>0$, $\omega^2_{\bf k}= {\bf k}^2+ m^2$, and $c_{1 \bf
k}$ and $c_{2,\bf k}$ are arbitrary constants of $O(|{\bf k}|^0)$.
For the stable modes, ${\bf k}^2 > -m^2$, only the positive
frequency modes are chosen to give time-independent $G_{\bf k}$,
\begin{eqnarray} \label{G}
G_{\bf k}= \left\{
 \begin{tabular}{ll}
 $\displaystyle \frac{1}{2\tilde{\omega}_{\bf k}}
\left[|c_{1\bf k}|^2e^{2 \tilde{\omega}_{\bf k} t} + |c_{2\bf
k}|^2e^{-2 \tilde{\omega}_{\bf k} t}
+ c_{1\bf k}c_{2 \bf k}^*+ c_{1\bf k}^*c_{2\bf k} \right],$ &
 ~~ $0\leq {\bf  k}^2< -m^2,$ \\
 $\displaystyle \frac{1}{2\omega_{\bf k}}$,& ~~
 ${\bf  k}^2 \geq -m^2  $. \\
 \end{tabular} \right.
\end{eqnarray}

In this paper, we follow the precarious renormalization
scheme~\cite{stevenson,pi1,jhyee}, which uses the coupling
constant renormalization condition,
\begin{eqnarray} \label{coupling:ren}
\frac{1}{\lambda} + \frac{1}{4}I_{-1}(m_R^2)= \frac{1}{\lambda_R}
.
\end{eqnarray}
From this, the definition of mass in Eq.~(\ref{m:qG}) becomes
\begin{eqnarray} \label{m:ren}
m^2(t)&=& \mu^2(t)+ \frac{\lambda}{2}\left[q^2(t)+ \int_{\bf
k}G_{\bf k}(t)\right] = \mu^2(t)+ \frac{\lambda}{2}\bar{q}^2(t) +
\frac{\lambda}{2}I_0(m^2),
\end{eqnarray}
where we explicitly denote the $t$ dependence in $m^2(t)$ for
later use and
\begin{eqnarray} \label{bq}
\bar{q}^2&=& q^2+\int_{\bf k}G_{\bf k}-I_0(m^2)
\end{eqnarray}
is a finite redefinition of variable $q$, which plays a crucial
role in the renormalization of the effective action. In the
present simple example, $\bar{q}^2= q^2 + \theta(-x) \int_{|{\bf
k}|<\bar{m} }G_{\bf k}$, where $\bar{q}=q$ for positive
mass-squared. Later in this paper, we write the effective action
and potential in terms of $\bar{q}$ since it is simpler.
Subtracting the equation at the renormalization point,
$(q_R,m_R>0)$, from Eq.~(\ref{m:ren}) and using the integral
reduction formula~(\ref{stevenson:fmr}) and (\ref{formula}) we get
\begin{eqnarray} \label{m:r}
m^2(t) -m_R^2 = \frac{\lambda_R}{2}\left[ \bar{q}^2- \bar{q}_R^2+
   \frac{m_R^2}{16\pi^2}\left(x \ln |x| -x +1\right) \right] .
\end{eqnarray}
This equation can be simplified by introducing dimensionless
variables, $\displaystyle \Phi^2=16\pi^2 \frac{\bar{q}^2}{m_R^2}$,
and $\displaystyle \kappa = -\frac{32\pi^2}{\lambda_R}$:
\begin{eqnarray} \label{x:Phi}
(1- \kappa)(x-1)- (\Phi^2-\Phi_R^2)= x \ln |x| .
\end{eqnarray}
The equation (\ref{x:Phi}) can be solved by graphical method,
i.e., by explicitly plotting each side of the equation as a
function of $x$.
\begin{figure}[htbp]
\begin{center}
\includegraphics[width=.6\linewidth,origin=tl]{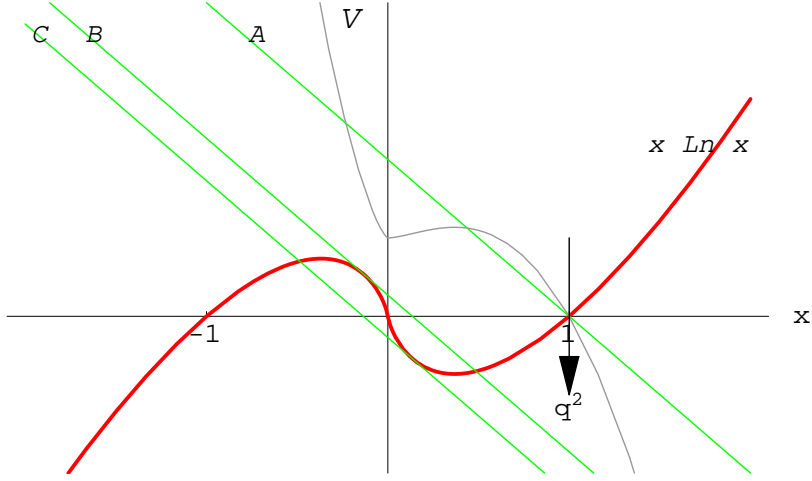}\hfill%
\end{center}
\caption{Graphical solution for Eq.~(\ref{x:Phi}). The thick curve
represents the function $x \ln |x|$, the thin curve is the
effective potential ${\cal V}$ at a given time, and the straight
lines A, B, C represent the left hand side of (\ref{x:Phi}) with
$\Phi^2-\Phi_R^2=0,~ \kappa-1-e^{-\kappa},~ \kappa -1 +
e^{-\kappa}$ respectively, for $\kappa >1$. The arrow indicates
how the lines move as $\Phi^2-\Phi_R^2$ increases. The potential
has a local minimum at $x=0$.} \label{qgt:fig}
\end{figure}
As an explicit example, we consider $\kappa >1$ case in Fig. 1.
The equation has three roots for $ -\kappa+1+e^{-\kappa}
<\Phi^2-\Phi_R^2 < -(-\kappa+1+e^{-\kappa})$, two roots for
$\Phi^2-\Phi_R^2=\pm(-\kappa+1+e^{-\kappa})$, and one root
otherwise. For $\Phi^2-\Phi_R^2=(\kappa -1)$, the straight line
passes through the origin. Note also that $\Phi^2-\Phi_R^2$ should
be greater than $-\Phi_R^2$. The improvements in this solution
from the previous works are the following: First, we always have
roots of Eq.~(\ref{x:Phi}) for any value of $\bar{q}^2\geq 0$.
Second, $x=0$ is not a special point now since the potential is
continued for negative $x$.

The divergent part of the effective Hamiltonian~(\ref{H}) consists
of two parts: $\displaystyle \frac{1}{2}\mu^2
q^2-\frac{\lambda}{8} \left(\int_{\bf k}\phi^*_{\bf k}\phi_{\bf
k}\right)^2$ and $\int_{\bf k}H_{G,\bf k}$. It is convenient to
change the first part using Eq.~(\ref{bq}) to give
\begin{eqnarray} \label{lll}
\frac{1}{2}\mu^2 q^2-\frac{\lambda}{8} \left(\int_{\bf k}G_{\bf
k}\right)^2=\frac{\mu^2}{2}\bar{q}^2+ \frac{m^2}{2}(q^2-\bar{q}^2)
-\frac{\lambda}{8}I_0^2(m^2)
+\frac{\lambda}{2}\bar{q}^2(q^2-\bar{q}^2),
\end{eqnarray}
where the divergences in $I_0^2$ and bare mass, $\mu^2$, are
separated from the finite contributions. If one incorporates the
renormalization condition~(\ref{coupling:ren}), the last term in
Eq.~(\ref{lll}) goes to zero since the bare coupling is $-0$. In
the second part, we separate the time-dependent part from
time-independent part using the explicit formula for $G_{\bf k}$,
Eq.~(\ref{G}),
\begin{eqnarray} \label{G:T}
\int_{\bf k}H_{G,\bf k}=\theta(-x) \bar{m}^4 V_L(\bar{m},t) +
\frac{1}{2}\int_{+,\bf k}\omega_{\bf k}=\theta(-x) \bar{m}^4
V_L(\bar{m},t)+ I_1(m^2),
\end{eqnarray}
where the time-dependent part is
\begin{eqnarray}
V_L(\bar{m},t)=-\frac{4}{\bar{m}^4} \int_{|{\bf
k}|<\bar{m}}(c_{1\bf k}c_{2\bf k}^*+c_{1\bf k}^*c_{2\bf
k})\tilde{\omega}_{\bf k}+ \frac{1}{\bar{m}^4}\int_{|{\bf
k}|<\bar{m}}\left[\frac{1}{8}+ (c_{1 \bf k}c^*_{2 \bf k}+ c_{1 \bf
k}^*c_{2 \bf k})^2\right]G_{\bf k}^{-1} .
\end{eqnarray}
Ignoring terms which go to zero when $\lambda \rightarrow -0$, we
get the effective Hamiltonian
\begin{eqnarray} \label{pot:eff}
H_{eff}(q,p)&=&\frac{p^2}{2}+ \frac{m^2}{2}(q^2-\bar{q}^2)+
\frac{\mu^2}{2}\bar{q}^2
 +I_1(m^2)-\frac{\lambda}{8}I_0^2(m^2)
  +\theta(-x) m_R^4 x^2V_L(\bar{m},t) \\
 &=&\frac{p^2}{2}+\frac{m^2}{2}(q^2-\bar{q}^2)
 + \theta(-x) m_R^4 x^2V_L(\bar{m},t)
 +V_D(\bar{q},m^2) ,   \nonumber
\end{eqnarray}
where the divergent part, $V_D$, becomes
\begin{eqnarray} \label{V:D}
V_D(\bar{q},m^2)&=&\frac{\mu^2}{2}\bar{q}^2+I_1(m^2)-\frac{\lambda}{8}I_0^2(m^2)
\\
&=& D+ \frac{\mu^2}{2}[\bar{q}^2-\bar{q}_R^2]+
I_1(m^2)-I_1(m_R^2)- \frac{\lambda}{8}[I_0(m^2)-I_0(m_R^2)]
\nonumber \\
&=&\left\{
  \begin{tabular}{ll}
$\displaystyle D+ \frac{1}{2}m_R^2 x(\bar{q}^2-\bar{q}_R^2)+
 \frac{m_R^4}{32\pi^2}L_3(x)-\frac{m_R^4}{2\lambda_R}(x-1)^2 $ ,
   &  ~~$ \displaystyle x=\frac{m^2}{m_R^2}> 0$ ,\\
  $\displaystyle D+\frac{1}{2}m_R^2x(\bar{q}^2-\bar{q}_R^2)
    + \frac{m_R^4}{32\pi^2}[L_3(-x)+2x]
      -\frac{m_R^4}{2\lambda_R}(x-1)^2  $ ,   & ~~
  $\displaystyle x=\frac{m^2}{m_R^2} <0$,
\end{tabular}
  \right. \nonumber
\end{eqnarray}
with $\displaystyle D= \frac{\mu^2}{2}\bar{q}_R^2+
I_1(m_R^2)-\frac{\lambda}{8} I_0^2(m_R^2)$ being a $q$ independent
divergent constant. Let us define the normalized effective
Hamiltonian by
\begin{eqnarray} \label{VD:nor}
{\cal H}&\equiv&\frac{H_{eff}-D}{m_R^4/(64\pi^2)} =64\pi^2
\theta(-x) x^2 V_L(\bar{m},t)+\frac{32\pi^2 p^2}{m_R^4}+
\frac{32\pi^2 q^2}{m_R^2}x - 2x \Phi_R^2 +2L_3(|x|) + 4 x
\theta(-x) -\kappa (x-1)^2 \\
&=&\frac{32\pi^2 p^2}{m_R^4}+  \theta(-x) \left[64\pi^2 x^2
V_L(\bar{m},t) -\frac{32\pi^2x}{m_R^2}\int_{|{\bf
k}|<\bar{m}}G_{\bf k} \right] + x(\Phi^2- \Phi_R^2)
-\frac{1}{2}(x-1)(x-1+2 \kappa)  \nonumber
   \\
&=&\frac{32\pi^2 p^2}{m_R^4}+ \theta(-x) \left[64\pi^2 x^2
V_L(\bar{m},t) -\frac{32\pi^2x}{m_R^2}\int_{|{\bf
k}|<\bar{m}}G_{\bf k} \right]- x^2 \ln |x|+ \left(
\frac{1}{2}-\kappa\right)(x^2-1) , \nonumber
\end{eqnarray}
where we write the Hamiltonian in several different forms using
Eq.~(\ref{x:Phi}).

Let us consider the case where the initial value is given by
$c_{2\bf k}=0$ and $c_{1 \bf k}=c_1$, to present an explicit form
of the effective potential. Then, $V_L(\bar{m},t)$ exponentially
decreases to zero as time $t\rightarrow \infty$ and
\begin{eqnarray} \label{G:small}
\int_{|{\bf k}|<\bar{m}}G_{\bf k} &=& \frac{|c_1|^2}{4 \pi^2}
\int_0^{\bar{m}} \frac{k^2 e^{2t \sqrt{\bar{m}^2-
k^2}}}{\sqrt{\bar{m}^2- k^2}} dk = \frac{|c_1|^2}{4 \pi^2}
\int_0^{\bar{m}}e^{f(k^2/\bar{m}^2)} dk
\end{eqnarray}
where the function $f$ is maximized at $\displaystyle {\bf
k}^2=\frac{\bar{m}}{t}$, and the integral can be performed using
the steepest descent method to give (See Appendix A):
\begin{eqnarray} \label{q:series}
\int_{|{\bf k}|<\bar{m}}G_{\bf k}\simeq \frac{|c_1|^2}{4 \pi^2}
\sqrt{\frac{\pi\bar{m}}{2t^3}} e^{2 \bar{m}t -1/2}.
\end{eqnarray}
Therefore, the renormalized effective potential at a given large
time $t$ becomes
\begin{eqnarray} \label{VD:nor2}
{\cal V}(x) &=& \left.{\cal H}\right|_{p=0}= 8 |c_1|^2
x^2g(m_R(-x)^{1/2} t)\theta(-x)- x^2 \ln |x|+ \left(
\frac{1}{2}-\kappa \right)(x^2-1) ,
\end{eqnarray}
where $\displaystyle g(y)=\sqrt{\frac{\pi}{y^3}} e^{2 y -1/2}$.
The renormalized potential is symmetric about $x=0$ if $c_1=0$.
The first term of the effective potential~(\ref{VD:nor2})
exponentially increases as time so that one cannot have a large
negative mass-squared at a later time.
\begin{figure}[tbhp]
\begin{center}
\includegraphics[width=.5\linewidth,origin=tl]{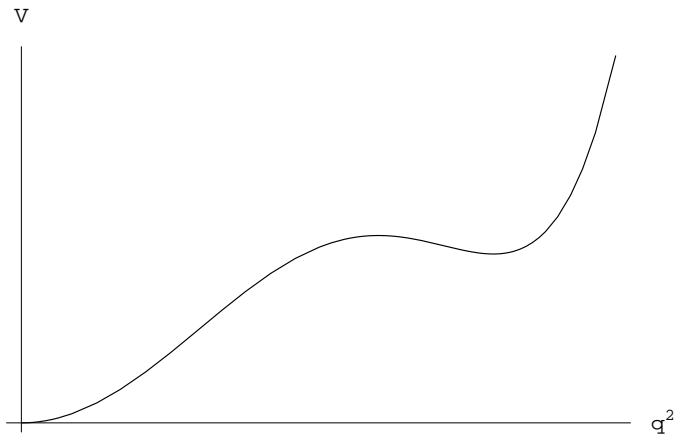}\hfill%
\end{center}
\caption{Typical form of the effective potential ${\cal V}$ at a
given time $t$. The horizontal axis represents $\bar{q}$. The
global minimum at the origin is at $q=0$ and the local minima at
$\bar{q} \neq 0$ is the point of $\bar{q}$ which corresponds to
$x=0$.} \label{pot1:fig}
\end{figure}

Now we discuss the shape of the effective potential as a function
of $\bar{q}$ as seen in Fig. 2. Since $\bar{q}=q$ for $m^2>0$,
this gives the effective potential for $q$ for positive
mass-squared case. As can be seen in Fig. 1, there is at least one
root $x$ for a given $\bar{q}$. In the case of two or three $x$
roots, the true $x$ values should be chosen such that it minimize
the effective potential~(\ref{VD:nor2}). Then the absolute minimum
of the potential is at $\Phi^2=0$. The potential increases as a
function of $\Phi^2$ until $\Phi^2-\Phi_R^2=e^{-\kappa}-1+\kappa$
where $x=e^{-\kappa}$. It then decreases until
$\Phi^2-\Phi_R^2=\kappa -1$ where $x=0$, which is a local minima
of the effective potential, and then it increases again. The
metastable state corresponds to $x=0$ so the theory is massless at
that point. The effect of the inclusion of the negative
mass-squared modes becomes apparent for large $\bar{q}^2$, in
which, the potential exponentially increases in time and the
mass-squared is negative. This means, for a realistic system, that
the modes with negative mass squared soon decays to positive
stable modes since it needs infinite energy to keep the negative
mode continually.

Some comments are in order. Due to the inclusion of negative modes
the solutions to the equation~(\ref{x:Phi}) always exist.
Stevenson~\cite{stevenson} found that the present technique is
inadequate to find the effective potential for negative $\kappa$
for the following reason: At $x=1$, we should have
$(m^2,q^2)=(m_R^2,q_R^2)$. There are two solution for $x$ to the
Eq.~(\ref{x:Phi}) if $\kappa <0$ and the solution at $x=1$
corresponds to the larger potential value of the two. This does
not obey the guiding principle which chooses the $x$ which gives
lower potential value. To overcome this difficulty he used a
rescaling of variables which maps $\kappa <0$ into $0<\kappa <1$.
This is also the case for the example above.

\section{Spinodal instability and finite smooth
quench by controlling the mass-squared}
In this section, we consider a finite smooth quench model, which
was previously considered in Ref.~\cite{spkim}. The finite quench
transition is described by a scalar field with the mass given by
\begin{eqnarray} \label{m:t}
m^2(t)= \frac{m_i^2-m_f^2}{2}-\frac{m_i^2+m_f^2}{2} \tanh
\left(\frac{t}{\tau}\right) ,
\end{eqnarray}
where $m_i^2$ and $m_f^2$ are both positive definite. At earlier
time, $t=-\infty$, the mass-squared has the initial value $m_i^2$
and at later time, $t=\infty$, the final value $-m_f^2$. Here
$\tau$ measures the quench rate: the large $\tau$ implies that the
mass changes slowly, whereas the small $\tau$ implies a rapid
change of the mass-squared. In particular, the $\tau\rightarrow 0$
limit corresponds to the instantaneous change from $m_i^2$ to
$-m_f^2$ at $t=0$. To find the Fock space for each mode one needs
to solve the classical equation of motion
\begin{eqnarray} \label{eom:ctrl}
\ddot{\phi}_{\bf k}(t) + \left[{\bf k}^2+
\frac{m_i^2-m_f^2}{2}-\frac{m_i^2+m_f^2}{2} \tanh
\left(\frac{t}{\tau}\right) \right] \phi_{\bf k}(t)=0 .
\end{eqnarray}
It should be noted that the long wavelength modes, $|{\bf k}| \leq
m_f$, lead to the change in the sign of the frequency at later
times $(t \gg \tau)$,
\begin{eqnarray} \label{tau}
\omega_{\bf k}^2(t)= {\bf k}^2- m_f^2 <0,
\end{eqnarray}
and suffer from spinodal instability.  Each long wavelength mode
has a different quench time determined by $\omega_{\bf k}^2(t_{\bf
k}) =0$. The solution to Eq.~(\ref{eom:ctrl}) are found separately
for stable modes and unstable modes. The stable modes $(k \geq
m_f)$ have the solution
\begin{eqnarray} \label{sol:S}
\phi_{\bf k}(t)= \frac{1}{\sqrt{2 \omega_{i, \bf k}}} e^ {-i
\omega_{i , \bf k}t} ~_2F_1(-i \tau \omega_{+,\bf k}, -i \tau
\omega_{-,\bf k}; 1-i \tau \omega_{i, \bf k} ; -e^{2t/\tau}),
\end{eqnarray}
where
\begin{eqnarray} \label{om:pm}
\omega_{\pm, \bf k} = \frac{\omega_{i, \bf k} \pm \omega_{f, \bf
k} }{2} ,
\end{eqnarray}
with
\begin{eqnarray} \label{om:if}
\omega_{i, \bf k}= \sqrt{{\bf k}^2+ m_i^2}, ~~\omega_{f, \bf k}=
\sqrt{{\bf k}^2- m_f^2}.
\end{eqnarray}
Whereas the unstable modes $(|{\bf k}|< m_f)$ have the solution
\begin{eqnarray} \label{sol:U}
\phi_{\bf k}(t)= \frac{1}{\sqrt{2 \omega_{i, \bf k}}} e^ {-i
\omega_{i , \bf k}t} ~_2F_1( \tau w_{\bf k},- \tau w_{\bf k}^*;
1-i \tau \omega_{i, \bf k} ; -e^{2t/\tau}),
\end{eqnarray}
where the complex frequency is given by
\begin{eqnarray} \label{om:pm}
w_{\bf k}=  \frac{\tilde{\omega}_{f, \bf k} + i \omega_{i, \bf k}
}{2} .
\end{eqnarray}

At earlier time $(t \ll -\tau)$ before the quench begins, both
solutions (\ref{sol:S}) and (\ref{sol:U}) have the same asymptotic
form
\begin{eqnarray} \label{initial}
\phi_{i, \bf k}(t)= \frac{1}{\sqrt{2 \omega_{i, \bf k}}}e^{-i
\omega_{i , \bf k}t} ,
\end{eqnarray}
which matches the initial condition~(\ref{phi:i}). At later time
$(t\gg \tau)$ after the completion of quench, we find, by using
the asymptotic form of the hypergeometric function~\cite{GR}, the
asymptotic form for the stable modes~(\ref{sol:S})
\begin{eqnarray} \label{S:as}
\phi_{f_S,{\bf k}}(t)&=& C_{S+,\bf k}\frac{e^{-i \omega_{f, \bf k}
t}}{\sqrt{2 \omega_{f, \bf k}}} + C_{S-,\bf k}\frac{e^{i
\omega_{f, \bf k} t}}{\sqrt{2 \omega_{f, \bf k}}},
\end{eqnarray}
where
\begin{eqnarray} \label{CSpm}
C_{S\pm,\bf k}= \sqrt{\frac{\omega_{f,\bf k}}{\omega_{i,\bf k}}}
    \frac{  \Gamma(1-i\omega_{i, \bf k}\tau)\Gamma(\mp i
    \omega_{f,\bf k} \tau) }{ \Gamma(1-i \omega_{\pm,\bf k} \tau)
    \Gamma(-i \omega_{\pm,\bf k} \tau) } ,
\end{eqnarray}
and the asymptotic form for the unstable modes~(\ref{sol:U})
\begin{eqnarray} \label{U:as}
\phi_{f_U,{\bf k}}(t)&=& C_{U+,\bf k} \frac{e^{ \tilde{\omega}_{f,
\bf k} t}}{\sqrt{2 \tilde{\omega}_{f, \bf k}}} + C_{U-,\bf
k}\frac{e^{- \tilde{\omega}_{f, \bf k} t}}{\sqrt{2
\tilde{\omega}_{f, \bf k}}},
\end{eqnarray}
where
\begin{eqnarray} \label{CUpm}
C_{U+,\bf k}= \sqrt{\frac{\tilde{\omega}_{f,\bf k}}{ \omega_{i,
\bf k}}}~
    \frac{  \Gamma(1-i\omega_{i, \bf k}\tau)\Gamma(
   \tilde{\omega}_{f,\bf k}\tau)  }{ w_{\bf k}^* \tau
    \Gamma^2( w_{\bf k}^* \tau)  }, ~~
C_{U-,\bf k}= \sqrt{\frac{\tilde{\omega}_{f,\bf k}}{ \omega_{i,
\bf k}}}~
    \frac{  \Gamma(1-i\omega_{i, \bf k}\tau)\Gamma(-
   \tilde{\omega}_{f,\bf k}\tau)  }{ -w_{\bf k} \tau
    \Gamma^2( -w_{\bf k} \tau)  }.
\end{eqnarray}

From these results, the transition rate from the initial vacuum to
the final vacuum and the coefficient of the negative frequency
solution leading to the particle production rate were calculated
in Ref.~\cite{spkim}. The asymptotic form for $\phi_{fU,\bf k}^*
\phi_{fU,\bf k}$ was also calculated to obtain the scaling
relation of the domain size
\begin{eqnarray} \label{scaling}
\xi_D(t) = \sqrt{\frac{8 \tilde{t}}{m_f}},
\end{eqnarray}
where
\begin{eqnarray} \label{tilde t}
\tilde{t}= t- \frac{\tau^3}{2} [\zeta(3)-1]|w_{\bf k}|^2 .
\end{eqnarray}

For use in the next section, we explicitly present some more
results not mentioned in Ref.~\cite{spkim}. The asymptotic form of
the absolute square of the stable modes are
\begin{eqnarray} \label{S:as}
G_{f_S,\bf k}=\phi_{f_S,\bf k}^*(t) \phi_{f_S,\bf k}(t) &=&
\frac{g^+_{\bf k}}{2 \omega_{f,\bf k}}\left[1 + R^+_{\bf k}
\cos\left(2\omega_{f,\bf k} t- \psi^+_{\bf k}\right) \right]  ,
\end{eqnarray}
with the time-independent constants $g_{\bf k}^+$, $R_{\bf k}^+$,
and $\psi_{\bf k}^+$ given by
\begin{eqnarray} \label{gR:S}
g^+_{\bf k}&=&\frac{\sinh^2 \pi \omega_{+,\bf k} \tau +\sinh^2\pi
\omega_{-,\bf k} \tau }{ \sinh \pi \omega_{i,\bf k}\tau \sinh \pi
\omega_{f,\bf k}\tau } \\
R^+_{\bf k} &=& \left[\frac{\sinh\pi \omega_{+,\bf
k}\tau}{\sinh\pi\omega_{-,\bf k}\tau}+\frac{\sinh\pi \omega_{-,\bf
k}\tau}{\sinh\pi\omega_{+,\bf k}\tau}\right]^{-1}, \\
i\psi^+_{\bf k}&=& \ln \frac{\Gamma(1-i\omega_{f,\bf
k}\tau)\Gamma(1+ i \omega_{+,\bf k}\tau)\Gamma(1-i\omega_{-,\bf
k}\tau)}{\Gamma(1+i\omega_{f,\bf k}\tau)\Gamma(1- i \omega_{+,\bf
k}\tau)\Gamma(1+i\omega_{-,\bf k}\tau)} ,
\end{eqnarray}
where $\psi_{\bf k}^+=0$ for $|{\bf k}| \ll 1/\tau$, and $g_{\bf
k}^+ =1$, $R_{\bf k}^+=0$ for $|{\bf k}|\gg 1/\tau$. On the other
hand, the ultra violet limit of $\phi_{f_S,\bf k}^*(t)
\phi_{f_S,\bf k}(t)$ is given by
\begin{eqnarray} \label{GS:large k}
G_{f_S, \bf k}(t) = \frac{1}{2 \omega_{f,\bf k}} + O(e^{-\pi
\omega_{i,\bf k}\tau} ) , ~~ {\bf k}^2 \gg \frac{1}{\tau^2} .
\end{eqnarray}
Using the properties of $\Gamma$ functions we get the
instantaneous quench limit ($\tau \rightarrow 0$) of the stable
modes and the unstable modes:
\begin{eqnarray} \label{phiU:tau=0}
\lim_{\tau \rightarrow 0}\phi_{f_S,\bf k}(t)&=& \frac{1}{2\sqrt{2
\omega_{f,\bf
  k}}}\left[ \left(\sqrt{\frac{\omega_{f,\bf k}}{\omega_{i,\bf
  k}}}+\sqrt{\frac{\omega_{i,\bf k}}{\omega_{f,\bf
  k}}}\right)e^{-i \omega_{f,\bf k}t}
+ \left(\sqrt{\frac{\omega_{f,\bf k}}{\omega_{i,\bf
k}}}-\sqrt{\frac{\omega_{i,\bf k}}{\omega_{f,\bf
k}}}\right)e^{i\omega_{f,\bf k}t} \right] , \\
\lim_{\tau \rightarrow 0}\phi_{f_U,\bf k}(t)&=& \frac{1}{2\sqrt{2
\omega_{f,\bf k}}}\left[ \left(\sqrt{\frac{\omega_{f,\bf
k}}{\omega_{i,\bf k}}}-i\sqrt{\frac{\omega_{i,\bf
k}}{\omega_{f,\bf k}}}\right)e^{\omega_{f,\bf k}t} +
\left(\sqrt{\frac{\omega_{f,\bf k}}{\omega_{i,\bf
k}}}+i\sqrt{\frac{\omega_{i,\bf k}}{\omega_{f,\bf
k}}}\right)e^{-\omega_{f,\bf k}t} \right] ,
\end{eqnarray}
and their absolute squares are
\begin{eqnarray} \label{GS:tau=0}
\lim_{\tau \rightarrow 0} G_{f_S, \bf k}(t)&=&
\frac{1}{2\omega_{f,\bf k}} \left( \Omega_{+,\bf k}-\Omega_{-,\bf
k} \cos 2 \omega_{f,\bf k} t \right), \\
\lim_{\tau \rightarrow 0} G_{f_U, \bf k}(t) &=&
\frac{1}{2\tilde{\omega}_{f,\bf k}} \left(\tilde{\Omega}_{-,\bf
k}+ \tilde{\Omega}_{+,\bf k} \cosh 2 \tilde{\omega}_{f,\bf k} t
\right) ,
\end{eqnarray}
where
\begin{eqnarray} \label{Omega:pm}
\Omega_{\pm,\bf k}=\frac{1}{2}\left(\frac{\omega_{f, \bf
k}}{\omega_{i, \bf k}}\pm \frac{\omega_{i, \bf k}}{\omega_{f, \bf
k}}\right), ~~\tilde{\Omega}_{\pm,\bf
k}=\frac{1}{2}\left(\frac{\tilde{\omega}_{f, \bf k}}{\omega_{i,
\bf k}}\pm \frac{\omega_{i, \bf k}}{\tilde{\omega}_{f, \bf
k}}\right).
\end{eqnarray}
All of these limiting properties are used in the next section to
obtain the WKB mode solutions with arbitrary time-dependent mass.

\section{Time evolution and renormalization of a system after
 instantaneous quenching}

The model described in Sec IV does not describe a real system
undergoing second order phase transition because the spinodal
instability increases indefinitely. This model describes an
intermediate process of the realistic phase transition toward the
spinodal line, $m^2(t_k)={\bf k}^2$, at which point the
instability of the mode $\phi_{\bf k}$ stops to increase. In this
section we consider a self-interacting scalar field model by
properly treating the back-reaction of the field through the
$\lambda \phi^4$ interaction. All the instabilities end to
increase at time ${\cal T}$ defined by $m^2({\cal T})=0$. Let the
mass-squared of the scalar field $m^2(t)$ be positive value
$m_i^2$ for $t<0$ and then it suddenly jumps to negative value
$-m_f^2$ at $t=\tau \ll m_i^{-1}$. We consider the quench to occur
in time $\tau$ which is far smaller than the normal time scale
($1/m$), which is why we use the word ``instantaneous" even though
it is not zero. And the time $\tau$ should not be zero since no
realistic physical processes can alter the mass-squared
instantaneously. Unlike the example in the previous section, we do
not keep the mass-squared, $m^2(t)$, fixed, instead we let the
mass-squared vary freely by self interaction after the quench
$t>\tau$. The self-interaction of the scalar field leads to the
gradually increasing mass-squared to a positive value as time.

The equation of motion for the mode solution~(\ref{ddphi:0}),
then, becomes
\begin{eqnarray} \label{ddphi:2}
\ddot{\phi}_{\bf k}(t) + [{\bf k}^2 + m^2(t)] \phi_{\bf k}(t)=0.
\end{eqnarray}
Since the mass-squared is negative at $t=\tau$, the modes for
$t\geq \tau$ are divided into two categories depending on the size
of momentum relative to the mass
\begin{eqnarray} \label{phi:kLS}
\phi_{\bf k}(t) = \left\{ \begin{tabular}{ll}
  $\displaystyle \phi_{\bf k}^U $,&  ~~~$ {\bf k}^2 \leq -m^2(t)$,
   \vspace{.1cm}\\
   $\displaystyle \phi_{\bf k}^S $, & ~~~${\bf k}^2 > -m^2(t)$. \\
\end{tabular} \right.
\end{eqnarray}
In the WKB approximation, the unstable mode, $\phi^U_{\bf k}$, has
exponential type solution and the stable mode, $\phi^S_{\bf k}$,
has oscillatory solution. Let us assume that $\omega_{\bf
k}^2(t)={\bf k}^2 + m^2(t)$ increases for $t>\tau$ due to the back
reactions. Then, the unstable mode, $\phi^U_{\bf k}$, becomes
stable at time $t_k$, given by ${\bf k}^2 + m^2(t_k)=0$.
Conversely, for a given time $t$ the modes $\phi^U_{\bf k}$ with
$|{\bf k}|^2\leq -m^2(t)$ are unstable and the modes $\phi^S_{\bf
k}$ with $|{\bf k}|^2 > -m^2(t)$ are stable.  Since the
mass-squared increases as time, the number of unstable modes
decreases and finally the unstable modes disappear at time ${\cal
T}$, where $m^2({\cal T})=0$, which means the ending of the
instability growth. Now, the dynamical process can be divided into
three cases depending on the regions of time. First is the time
before the transition $t<0$ which is given by the initial
condition, second is the spinodal development process, $ 0<t<
{\cal T}$, the phase transition, and the final one is the
stabilization process after the termination of the instability
growth, $t>{\cal T}$.

 Let us assume $q(t)=0$ and $m^2(t) =m_i^2 >0$ for $t<0$.
Then, the initial state becomes
\begin{eqnarray} \label{phi:0}
\phi_{\bf k}(t<0)= \frac{1}{\sqrt{2 \omega_{i,{\bf k}} }} e^{-i
\omega_{i, {\bf k}} t},
\end{eqnarray}
and Eq.~(\ref{omega:k}) yields the gap equation
\begin{eqnarray} \label{gap:2}
\omega_{i, {\bf k}}^2 =m_i^2+ {\bf k}^2= \mu_i^2+ {\bf k}^2 +
\frac{\lambda \hbar}{2} \int_{\bf k} \frac{1}{2 \omega_{i,{\bf
k}}} \coth \left(\frac{\beta \omega_{i,{\bf k}} }{2} \right) ,
\end{eqnarray}
where $\mu_i$ is the initial bare mass and $\lambda$ is the bare
coupling. The infinities in the ${\bf k}$ integral are absorbed
into the bare mass and the bare coupling leading to finite and
positive $m_i^2$. For convenience we use the notation
$\tilde{\omega}^2_{\bf k}(t)=- \omega^2_{\bf k}(t) \geq 0 $ for
the modes ${\bf k}^2 \leq -m^2(t)$;
\begin{eqnarray} \label{tildeOmega}
\tilde{\omega}_{\bf k}^2(t) =\bar{m}^2(t)-{\bf k}^2= -\mu^2(t)-
{\bf k}^2 - \frac{\lambda}{2} \int_k \phi_{\bf k}^*(t)\phi_{\bf
k}(t) \coth \left(\frac{\beta \omega_{i,{\bf k}} }{2} \right) \geq
0 .
\end{eqnarray}

Using Eq.~(\ref{phiU:tau=0}) and the WKB approximation of
Eq.~(\ref{ddphi:2}), we get the solution for a given time
$t>\tau$,
\begin{eqnarray} \label{WKB}
\phi_{\bf k}^U(t)&=&C_{U+,\bf k}\frac{e^{ \int_0^t
\tilde{\omega}_{\bf k}(t') dt'}}{\sqrt{2 \tilde{\omega}_{\bf k}(t)
}}+C_{U-,\bf k}\frac{e^{ -\int_0^t \tilde{\omega}_{\bf k}(t')
dt'}}{\sqrt{2 \tilde{\omega}_{\bf k}(t) }}, ~~
 ~~ 0\leq {\bf k}^2 < -m^2(t), \\
\phi_{\bf k}^S(t)&=& \left\{
\begin{tabular}{ll}
$\displaystyle C_{U+,\bf k}\phi_{U,\bf k}^+  -C_{U-,\bf k}
\phi_{U,\bf k}^-
 ,$& ~~ $ -m^2(t) \leq {\bf k}^2 < m_f^2,$ \vspace{.1cm} \\
$\displaystyle  C_{S+,\bf k}\frac{e^{-i \int_0^t \omega_{\bf
k}(t') dt'}}{\sqrt{2 \omega_{\bf k}(t) }}+C_{S-,\bf k}\frac{e^{
i\int_0^t \omega_{\bf k}(t') dt'}}{\sqrt{2 \omega_{\bf k}(t) }},$
& ~~$ ~~ {\bf k}^2 \geq m_f^2$,
\end{tabular} \right.
\end{eqnarray}
where we use the matching condition,
\begin{eqnarray} \label{wkb:t1}
\frac{1}{\sqrt{2 \tilde{\omega}_{\bf k}(t) }}e^{ -\int_t^{t_k}
\tilde{\omega}_{\bf k}(t') dt'}  ~&\longleftrightarrow & ~
\frac{\sqrt{2 \pi} {\rm Ai}[-\mu_k(t-t_k)]}{\mu_k^{1/2}}
~\longleftrightarrow \phi^{+}_{U,\bf k}=
  \sqrt{\frac{2}{ \omega_{\bf k}(t)}} \cos \left(\int_{t_k}^t
 \omega_{\bf k}(t')dt' -\frac{\pi}{4} \right),
\\
\frac{1}{\sqrt{2 \tilde{\omega}_{\bf k}(t) }}e^{\int_t^{t_k}
\tilde{\omega}_{\bf k}(t') dt'}  ~&\longleftrightarrow & ~
 \sqrt{\frac{\pi}{2}}\frac{ {\rm
Bi}[-\mu_k(t-t_k)]}{\mu_k^{1/2}} ~\longleftrightarrow
~\phi^{-}_{U,\bf k}=
  -\frac{1}{\sqrt{2 \omega_{\bf k}(t)}} \sin \left(\int_{t_k}^t
 \omega_{\bf k}(t')dt' -\frac{\pi}{4} \right), \nonumber
\end{eqnarray}
of WKB approximation at time $t=t_k$, where $\displaystyle
\mu_k=\left[\frac{dm^2}{dt}(t_k) \right]^{1/3}$, and ${\rm Ai}(x)$
and $ {\rm Bi}(x)$ are the Airy functions. Later in this paper, we
use the notation
\begin{eqnarray} \label{thetas}
\tilde{\theta}_{\bf k}(t)=\int_0^{t} \tilde{\omega}_{\bf k}(t')
dt',~~~ \theta^+_{\bf k}(t)= \int_{0}^t \omega_{\bf k}(t') dt',~~~
\theta^-_{\bf k}(t)=\int_{t_k}^t \omega_{\bf k}(t') dt' .
\end{eqnarray}

In the case of the unstable modes, $|{\bf k}|<\bar{m}(t)$, the
frequency satisfies $|{\bf k}| \tau \ll 1$, which leads to
\begin{eqnarray} \label{phi:L}
\phi_{\bf k}^U(t) &=& \frac{1}{2\sqrt{2 \tilde{\omega}_{\bf
  k}(t)}}\left[ \left(\sqrt{\frac{\tilde{\omega}_{f,\bf k}}{
      \omega_{i,\bf k}}}
  -i\sqrt{\frac{\omega_{i,\bf k}}{\tilde{\omega}_{f,\bf
  k}}}\right)e^{\tilde{\theta}_{\bf k}(t)}
+ \left(\sqrt{\frac{\tilde{\omega}_{f,\bf k}}{\omega_{i,\bf
 k}}}+i\sqrt{\frac{\omega_{i,\bf k}}{\tilde{\omega}_{f,\bf
 k}}}\right)e^{-\tilde{\theta}_{\bf k}(t)} \right].
\end{eqnarray}
Therefore the WKB extended stable mode $\phi^S_{\bf k}(t)$ for
$-m^2(t) < {\bf k}^2 \leq m_f^2 $ becomes
\begin{eqnarray} \label{phi:t}
\phi^S_{\bf k}(t)=
\frac{1}{2}\left(\sqrt{\frac{\tilde{\omega}_{f,\bf k}}{
      \omega_{i,\bf k}}}
  -i\sqrt{\frac{\omega_{i,\bf k}}{\tilde{\omega}_{f,\bf
  k}}}\right)e^{\tilde{\theta}_{\bf k}(t_k)} \phi^+_{U,\bf k}(t)
+ \frac{1}{2}\left(\sqrt{\frac{\tilde{\omega}_{f,\bf
k}}{\omega_{i,\bf
 k}}}+i\sqrt{\frac{\omega_{i,\bf k}}{\tilde{\omega}_{f,\bf
 k}}}\right) e^{-\tilde{\theta}_{\bf k}(t_k)} \phi^-_{U,\bf k}(t).
\end{eqnarray}
Similarly, the stable modes with moderate frequencies $(m_f\leq
|{\bf k}| \ll 1/\tau )$ become
\begin{eqnarray} \label{phi:S}
\phi_{\bf k}^S(t) &=& \frac{1}{2\sqrt{2 \omega_{\bf
  k}(t)}}\left[ \left(\sqrt{\frac{\omega_{f,\bf k}}{\omega_{i,\bf
  k}}}+\sqrt{\frac{\omega_{i,\bf k}}{\omega_{f,\bf
  k}}}\right)e^{-i \theta^+_{\bf k}(t)}
+ \left(\sqrt{\frac{\omega_{f,\bf k}}{\omega_{i,\bf
k}}}-\sqrt{\frac{\omega_{i,\bf k}}{\omega_{f,\bf
k}}}\right)e^{i\theta^+_{\bf k}(t)} \right],
\end{eqnarray}
and the stable modes with very high frequencies $(|{\bf k}| \gg
1/\tau)$ become
\begin{eqnarray} \label{phi:SUV}
\phi^S_{\bf k}(t) = \frac{\omega_{i,\bf k}^{-i \omega_{i, \bf k}
\tau} \omega_{f,\bf k}^{-i \omega_{f,\bf k} \tau} }{\omega_{+,\bf
k}^{-2i\omega_{+,\bf k} \tau}}\frac{e^{-i \theta^+_{\bf
k}(t)}}{\sqrt{2 \omega_{\bf k}(t)}}  + O(e^{-\pi \omega_{i,\bf k}
\tau}) .
\end{eqnarray}
As shown in Eqs.~(\ref{phi:S}) and (\ref{phi:SUV}) the UV behavior
of the stable modes takes different form depending on the value of
the large momentum compared to $1/\tau$.

The absolute square of this mode solution, $G_{\bf k}(t)=\phi_{\bf
k}^*(t)\phi_{\bf k}(t) $, can be written as
\begin{eqnarray} \label{GL1}
G^U_{\bf k}(t) &=& \frac{1}{2\tilde{\omega}_{\bf k}(t) } \left[
\tilde{\Omega}_{-,\bf k} +\tilde{\Omega}_{+,\bf k}\cosh
  2\tilde{\theta}_{\bf k}(t)\right], {\bf k}^2< -m^2(t), \\
G^S_{\bf k}(t) &=& \left\{
\begin{tabular}{ll}
$\displaystyle \frac{g^-_{\bf k}}{2\omega_{\bf k}(t) } \left[1+
R^-_{\bf k}\cos\left(2\theta^-_{\bf k}(t)- \psi^-_{\bf k}\right)
\right]
 $, &~~$-m^2(t) < {\bf k}^2 \leq m_f^2 $ \\
$ \displaystyle \frac{g^+_{\bf k}}{2\omega_{\bf k}(t) } \left[1+
R^+_{\bf k}\cos\left(2\theta^+_{\bf k}(t)- \psi^+_{\bf k}\right)
\right]
 $& ,~~${\bf k}^2 \geq m_f^2 $, \\
 \end{tabular} \right.
\end{eqnarray}
where $g^+_{\bf k}$, $R^+_{\bf k}$, and $\theta^+_{\bf k}$ are
given in Eq.~(\ref{gR:S}), and
\begin{eqnarray} \label{gR:U}
g^-_{\bf k}&=&\tilde{\Omega}_{+,\bf k}
  \left[e^{2\tilde{\theta}_{\bf k}(t_k)}
  + \frac{1}{4}e^{-2\tilde{\theta}_{\bf k}(t_k)} \right], \\
R^-_{\bf k}&=& \left[\left(\frac{e^{4\tilde{\theta}_{\bf k}(t_k)}-
  1/4}{e^{4\tilde{\theta}_{\bf k}(t_k)}+
  1/4}\right)^2+ \left(\frac{\tilde{\Omega}_{-,\bf
  k}}{\tilde{\Omega}_{+,\bf k}}\right)^2
  \frac{16}{\left(4e^{2\tilde{\theta}_{\bf k}(t_k)}
   + e^{-2\tilde{\theta}_{\bf k}(t_k)}
  \right)^2} \right]^{1/2}, \\
\tan\psi^-_{\bf k}&=&\frac{\tilde{\Omega}_{+,\bf
  k}}{\tilde{\Omega}_{-,\bf k}}
  \left[e^{2\tilde{\theta}_{\bf k}(t_k)}
  - \frac{1}{4}e^{-2\tilde{\theta}_{\bf k}(t_k)} \right] .
\end{eqnarray}
Note also that $G^S_{\bf k}(t)$ in the UV limit ($|{\bf k}|\gg
1/\tau)$ is
\begin{eqnarray} \label{GS:UV}
{\phi_{\bf k}^S}^*(t)\phi_{\bf k}^S(t)=\frac{1}{2 \omega_{\bf
k}(t)} + O(e^{-\pi \omega_{i,\bf k} \tau }).
\end{eqnarray}
On the other hand, in the intermediate region ($m_f\leq |{\bf k}|
\ll 1/\tau$), it becomes
\begin{eqnarray} \label{G:S}
{\phi_{\bf k}^S}^*(t)\phi_{\bf k}^S(t) = G_{\bf k}^M(t)\equiv
\frac{1}{2 \omega_{\bf k}(t)}\left[ \Omega_{+,\bf k}+
\Omega_{-,\bf k}\cos 2\theta^+_{\bf k}(t) \right], ~~ m_f\leq
|{\bf k}| \ll 1/\tau .
\end{eqnarray}
Note also that the time derivatives of $G$'s satisfy
\begin{eqnarray} \label{dG}
\dot{G}^U_{\bf k}(t) &=& \frac{\dot{m^2}}{2
  \tilde{\omega}_{\bf k}^2} G^{U}_{\bf k}(t)
   +\tilde{\Omega}_{+,\bf k} \sinh 2 \tilde{\theta}_{\bf k}(t), \\
 \dot{G}^S_{\bf k}(t) &=& - \frac{\dot{
m^2}(t)}{2 \omega^2_{\bf k}(t)} G^S_{\bf k}(t)- g^\pm_{\bf k}
R^\pm_{\bf k}\sin\left[2 \theta^\pm_{\bf k}(t)-\psi^\pm_{\bf
k}\right].
\end{eqnarray}
Therefore, the $H_{\bf k}$ in Eq.~(\ref{Hk}) becomes
\begin{eqnarray} \label{GdG^2}
H^S_{\bf k}&=& \frac{(\dot{m^2})^2}{32 \omega_{\bf k}^4} G_{\bf
k}^S + \frac{\dot{m^2}}{8\omega_{\bf k}^2}g^\pm_{\bf k} R^\pm_{\bf
k} \sin \left[2\theta^\pm_{\bf k}(t)-\psi^\pm_{\bf k}\right]+
\frac{1}{2} g_{\bf k}^\pm \omega_{\bf k} + \frac{1}{8}
\left[1-g_{\bf k}^{\pm 2}(1- R_{\bf
k}^{\pm 2} )\right] (G_{\bf k}^S)^{-1} ,  \\
  \label{HU}
H^U_{\bf k}&=&  \frac{(\dot{m^2})^2}{32 \tilde{\omega}_{\bf k}^4}
G_{\bf k}^U + \frac{\dot{m^2}}{8\tilde{\omega}_{\bf
k}^2}\tilde{\Omega}_{+,\bf k} \sinh 2\tilde{\theta}_{\bf k}(t)+
\tilde{\omega}_{\bf k}^2 G^U_{\bf k}
-\frac{1}{2}\tilde{\omega}_{\bf k}\tilde{\Omega}_{-,\bf k} .
\nonumber
\end{eqnarray}
where we can ignore the last term of (\ref{GdG^2}) for ${\bf k}^2
\geq m_f^2$ since $\displaystyle g_{\bf k}^{+2}(1- R_{\bf k}^{+2}
)= \left[\frac{\sinh^2\pi \omega_{+,\bf k}
\tau-\sinh^2\pi\omega_{-,\bf k}\tau}{\sinh \pi\omega_{f,\bf k}
\tau \sinh \pi \omega_{i,\bf k}\tau }\right]^2 \simeq 1$ for most
range of $|{\bf k}|> \bar{m}$.

The ultra-violet divergences are related only to $\phi_{\bf k}^S$
modes. From the structure of $G^S_{\bf k}$ written in Eqs.
(\ref{GS:UV}) we see that the only divergence in $G^S_{\bf k}$
comes from $1/[2\omega_{\bf k}(t)]$ term. It is already shown that
the equation of motion for this form of $G_{\bf k}$ is UV
renormalizable~\cite{pi1,kim} with the condition given
in~(\ref{coupling:ren}). Using Eqs.~(\ref{lll}) and (\ref{GdG^2}),
we get the effective Hamiltonian
\begin{eqnarray} \label{H:ren}
H= H_{f}+ \frac{1}{2}\mu^2\bar{q}^2+
I_1(m^2)-\frac{\lambda}{8}I_0^2(m) + \frac{1}{2}\int_{+,\bf k}
(g_{\bf k}^\pm-1) \omega_{\bf k}
\end{eqnarray}
where
\begin{eqnarray} \label{Hfin}
H_{f}&=&\frac{1}{2}\dot{q}^2+ \frac{1}{2}
m^2(t)[q^2(t)-\bar{q}^2(t)]+
\frac{\lambda}{2}\bar{q}^2(q^2-\bar{q}^2)+\frac{\lambda}{4!}q^4
+(\dot{m^2})^2 \int_{{\bf k}} \frac{G_{\bf k}}{32 \omega_{\bf
k}^4} \\
&+&\frac{\dot{m^2}}{8} \left\{ \int_{+,{\bf k}} \frac{g_{\bf
k}^\pm R_{\bf k}^\pm}{\omega_{\bf k}^2} \sin \left[2\theta^+_{\bf
k}(t)-\psi_{\bf k}^\pm\right]+\int_{{\bf k}^2<-m^2}
\frac{\tilde{\Omega}_{+,\bf k}}{\tilde{\omega}_{\bf k}^2} \sinh
2\tilde{\theta}_{\bf k}(t)\right\}+\int_{{\bf
k}^2<-m^2}\tilde{\omega}_{\bf k}^2 G^U_{\bf k}-
\frac{1}{2}\int_{{\bf
k}^2<-m^2} \tilde{\omega}_{\bf k}\tilde{\Omega}_{-,\bf k} \nonumber \\
&+& \frac{1}{8} \int_{+,\bf k}\left[1-g_{\bf k}^{\pm 2}(1- R_{\bf
k}^{\pm 2} )\right] (G_{\bf k}^S)^{-1} . \nonumber
\end{eqnarray}
The $\pm$ sign in $g^\pm_{\bf k}$, $R^\pm_{\bf k}$, and
$\theta_{\bf k}^\pm$ should be chosen by the sign of $|{\bf k}|
-m_f$.  The large momentum expansion of $g_{\bf k}^+ -1$ becomes
\begin{eqnarray} \label{intermediate}
g_{\bf k}^+ -1= \left\{
 \begin{tabular}{ll}
 $\displaystyle \frac{(\omega_{f,\bf k}-\omega_{i,\bf k})^2}{2
 \omega_{i,\bf k} \omega_{f,\bf k}}$,&~~ $ \bar{m}
    \leq |{\bf k}| \ll 1/\tau$,\vspace{.1cm}\\
 $\displaystyle  \frac{(m_i^2+ m_f^2)^2}{8 \omega_{\bf
k}^4(t)}+
\cdots,$ & ~~ $\bar{m} \ll |{\bf k}| \ll 1/\tau$, \vspace{.1cm}\\
$0$, &~~ $|{\bf k}| \gg 1/\tau$ ,\\
\end{tabular} \right.
\end{eqnarray}
which gives a large contribution proportional to $\ln (m_R\tau)$
to the integral. This logarithmic contribution to the energy,
which is related to the ``instantaneous" quench process, changes
the vacuum structure of the system in the sense that the
transition probability from the initial vacuum to the final vacuum
goes to zero in the $\tau \rightarrow 0$ limit. Since such
zero-probability transition is not physical, we should restrict
$\tau \neq 0$ so that the transition probability between the two
vacuum does not vanish. Let us define the integral,
\begin{eqnarray} \label{addI}
\bar{I}_{-1}(m^2,m_f) &\equiv& \int_{m_f}^{1/\tau} \frac{k^2 dk}{4
\pi^2 \omega_{\bf k}^3} \\
&=&\frac{1}{4\pi^2}\left[-\ln m_f\tau
-(1+\bar{m}^2\tau^2)^{-\frac{1}{2}}
+\left(1+\frac{\bar{m}^2}{m_f^2}\right)^{-1/2}
+\ln\frac{1+\sqrt{1+\bar{m}^2\tau}}{1+
\sqrt{1+\bar{m}^2/m_f^2}}\right],
 \nonumber\\
&=&\frac{1}{4\pi^2}\left[-\ln m_f \tau -
1+\left(1+|x|/x_f\right)^{-1/2} -\ln \frac{1+
\sqrt{1+|x|/x_f}}{2}\right] ,
 \nonumber
\end{eqnarray}
where $x_f=m_f^2/m_R^2$, we ignored terms which vanish in the
$\tau \rightarrow 0$ limit in the last equality, and we defined
the integration range from $m_f$ to avoid the infra-red (IR)
divergence which appear at the $k$ with $\omega_{\bf k}(t)=0$ if
$m^2(t)<0$. We denote the finite part of the last integral in
(\ref{H:ren}) by
\begin{eqnarray}
V_{\tau}=\frac{1}{2}\int_{+,\bf k} (g_{\bf k}^+ -1)\omega_{\bf
k}(t)-
 \frac{(m_i^2+m_f^2)^2}{8}\bar{I}_{-1}(m^2,m_f).
\end{eqnarray}
Then, the Hamiltonian becomes
\begin{eqnarray} \label{H:ren2}
H=H_{f}+ V_{\tau}+ \frac{1}{2}\mu^2 \bar{q}^2
+I_1(m)-\frac{\lambda}{8}I_0^2(m) +
\frac{(m_i^2+m_f^2)^2}{8}\bar{I}_{-1}(m^2,m_f),
\end{eqnarray}
where $H_f+V_{\tau}$ is the UV finite part of the Hamiltonian.
Since the form of the divergences of the Hamiltonian is the same
as Eq.~(\ref{pot:eff}), the renormalization can similarly be done:
\begin{eqnarray} \label{Hren:3}
H= D+H_{f} +V_{\tau}+ \frac{m_R^4}{64\pi^2}\left[
x(\Phi^2-\Phi_R^2)-\frac{1}{2}(x-1)(x-1+ 2 \kappa)\right]
 +\frac{(m_i^2+ m_f^2)^2}{8} \bar{I}_{-1}(m^2,m_f),
\end{eqnarray}
where the divergent constant, $\displaystyle D=\frac{1}{2} \mu^2
\bar{q}_R^2+ I_1(m_R)+ \frac{\lambda}{8}I_0^2(m_R)$. The last term
in this Hamiltonian contains a large constant contribution, which
is the energy added by the instantaneous quench at $t=0$.

If an observer is confined for $t>\tau$ and the experiments is
done with the energy scale ${\bf k}^2 \ll 1/\tau^2$, the
experiments cannot probe the existence of the very high frequency
behavior~(\ref{GS:UV}) of $G_{\bf k}$. In this case, one may do
additional renormalization of the $\ln m_R\tau$ term which is
related to the sudden change of the mass-squared at $t=0$,
\begin{eqnarray} \label{H:NewRen}
H&=& D'+H_{f} +V_{\tau}+ \frac{m_R^4}{64\pi^2} {\cal V}_{vac}, \\
{\cal V}_{vac}&=& x(\Phi^2-\Phi_R^2)-\frac{1}{2}(x-1)(x-1+ 2
\kappa) + 2(x_i+ x_f)^2\left[- 1+\left(1+|x|/x_f\right)^{-1/2}
-\ln \frac{1+ \sqrt{1+|x|/x_f}}{2} \right]  ,
\end{eqnarray}
where $x_i=m_i^2/m_R^2$, $x_f=m_f^2/m_R^2$, and   $\displaystyle
D'= D-\frac{(m_i^2+ m_f^2)^2}{32 \pi^2}\ln m_f \tau$. The
divergence in $\bar{I}_{-1}(m^2,m_f)$ is the origin of the
time-dependent renormalization, Eqs.~(5.8) and (5.9) of
Ref.~\cite{boyanovsky}. In some sense, ${\cal V}_{vac}$ denotes
the vacuum structure since it comes from the UV behaviors of the
mode solutions which will not be affected by small excitations of
moderate modes. In this point of view, $H_f+V_\tau$ may be
considered as an excitation to the vacuum state. It may be helpful
to figure out the structure of the potential ${\cal V}_{vac}$. The
new term ${\cal V}_{vac}- {\cal V}$ always decreases in $x$, which
leads to the possibility of symmetry breaking. Since the
mass-squared satisfies Eq.~(\ref{m:ren}), the equation which
determines $x$ by $\Phi$ is given by Eq.~(\ref{x:Phi}) and its
solution is obtained by graphical method in Fig. 1.

Note also that in Eq.~(\ref{bq}) the difference,
\begin{eqnarray} \label{bq-q}
\bar{q}^2-q^2= \theta(-x)\int_{|{\bf k}|<\bar{m}}G^U_{\bf k}
+\int_{+,\bf k}\frac{1}{2\omega_{\bf k}} \left[g_{\bf k}^{\pm}-1 +
g_{\bf k}^\pm R_{\bf k}^\pm \cos\left(2 \theta^\pm_{\bf
k}(t)-\psi_{\bf k}^\pm \right) \right]
\end{eqnarray}
is positive definite on time average due to
Eq.~(\ref{intermediate}). Therefore, the minimum value of
$\bar{q}^2$ should be a positive number.

Note that the integral $\int_{\bf k}G_{\bf k}/\omega_{\bf k}^4$ in
Eq.~(\ref{Hfin}) is apparently IR divergent at the value of ${\bf
k}$ where ${\bf k}^2+m^2=0$ if $m^2<0$. As one may see in the
previous literatures~\cite{pi1,kim}, there is no IR divergence in
the equations of motion expressed in $q$ and $G_{\bf k}$.
Therefore, this IR divergences come from the approximation of
$G_{\bf k}$ using the WKB solution. This origin of the IR
divergence reminds us the validity range of the WKB approximation,
\begin{eqnarray} \label{WKB:x}
 \left|\frac{\hbar\dot{m^2}(t)}{2\omega^2(t)}\right|
\ll 1 ,
\end{eqnarray}
which clearly fails to be satisfied at ``the classical point"
$t=t_k$ where $\omega_{\bf k}^2(t_k)=0$. In this sense, the IR
divergence is the artifact of the WKB approximation around the
transition point of $\omega_{\bf k}^2(t)$. The IR divergence does
not have physical origin but comes from the bad choice of the
solution of $\phi_{\bf k}$ near $t=t_k$. As an evidence of this
argument, if one uses the exact solutions (Airy functions) near
$t=t_k$, no divergences appear. To remedy this IR divergence, we
need a generalized WKB approximation~\cite{robicheaux}, which is
beyond the subject of this paper.

Fortunately and in fact naturally because this IR divergence comes
from the failure of the WKB solution to satisfy the validity range
of the WKB approximation~(\ref{WKB:x}), this IR divergences
present only in the kinetic terms for $m^2(t)<0$, which do not
influence the effective potential at a given time. The equations
of motion which determine the $G_{\bf k}$ functions are also IR
finite, which is related the physical correlation length.
Moreover, the effective action after the termination of the
spinodal transition, $t>{\cal T}$, is also IR finite. In the
present paper we restrict our interest only to these cases.

\section{Large instability approximation}
In the last section, we renormalized the effective action and
potential. In this section, we calculate the renormalized
effective Hamiltonian and potential for the case ${\cal T} \gg
1/m$. The validity of this approximation should be checked by the
dynamics at the time $\tau \leq t \leq {\cal T}$, which needs the
full understanding of the IR properties mentioned at the end of
the last section. In the absence of this knowledge, we can simply
assume $m_f$ to be large and the renormalized coupling to be small
so that it takes long enough time to increase the mass-squared to
zero from $-m_f^2$. This approximation leads to the unstable modes
to have large instability because of the exponential increase of
the WKB solution, which is the reason we call it the large
instability approximation.

Most part of this section is devoted to the evolution of the
system at time $t> {\cal T}$ except for the short calculation of
the effective potential for $m^2<0$ at the end of the this
section. Therefore, the mass-squared is always positive definite
and there is no unstable modes. The $G^S_{|{\bf k}|<m_f}$
function, in this approximation, is dominated by the exponentially
growing term given by
\begin{eqnarray} \label{GS:Large t}
G^S_{\bf k}(t)= \frac{\tilde{\Omega}_{+,\bf k}}{2 \omega_{\bf
k}(t)}e^{2 \int_0^{t_k} \tilde{\omega}_{\bf k}(t') dt'} \left[1+
\sin 2\theta^-_{\bf k}(t) \right], ~~ {\bf k}^2 < m_f^2,
\end{eqnarray}
and its integral over ${\bf k}$,
\begin{eqnarray} \label{intGs}
\int_{|{\bf k}|<m_f^2} G^S_{\bf k}(t) = \frac{1}{2\pi^2}
\int_0^{m_f} dk \frac{\tilde{\Omega}_{+,\bf k}}{2 \omega_{\bf
k}(t)}\left[1+ \sin 2\theta^-_{\bf k}(t) \right] e^{f(k^2/m_f^2)}
,
\end{eqnarray}
is approximately integrated by using the steepest descent method
in appendix B. The resulting integral formula are summarized in
Eqs.~(\ref{g:approx}) and (\ref{approx2}). Before proceeding
further, we define some parameters
\begin{eqnarray} \label{bl}
\bar{l}&=& \left(\frac{m_f}{2\pi \alpha\bar{t}}\right)^{3/2}e^{2
\tilde{\theta}_{\bar{k}}(\bar{t})}, \\
\bar{k}&=&\left(\frac{m_f}{\alpha\bar{t}}\right)^{1/2} , \\
\kappa &=& \bar{k} + i \bar{k}\Theta(t),
\end{eqnarray}
where $\bar{l}$ is a large scale which determines the instability,
$1\leq \alpha \leq 2$ is a number, and $\bar{t}$ is a time scale
determined by $\bar{k}^2+ m^2(\bar{t})=0$, both of which are
implicitly dependent on $\bar{k}$ and $\Theta(t)=\Theta(t,r=0)$
defined in Eq.~(\ref{Theta}).

We can approximate the integral in Eq.~(\ref{intGs}) as
\begin{eqnarray} \label{GL:longtime}
\int_{|{\bf k}|<m_f} G^S_{\bf k}(t) &\simeq &
  \bar{l}\left\{ \frac{\tilde{\Omega}_{+,\bar{k}}}{
  2\omega_{\bar{k}}(t)}+ \frac{e^{-2 \Theta^2(t)}}{4 i}
   \left[\frac{\tilde{\Omega}_{+,\kappa}}{
     2\omega_{\kappa}(t)}e^{2 i \theta^-_{\bar{k}}(t)}-
     \frac{\tilde{\Omega}_{+,\kappa^*}}{
     2\omega_{\kappa^*}(t)}e^{-2 i \theta^-_{\bar{k}}(t)}
  \right]\right\} .
\end{eqnarray}
Therefore,
\begin{eqnarray} \label{bq:q2}
\bar{q}^2-q^2 \simeq \bar{l}\left\{
\frac{\tilde{\Omega}_{+,\bar{k}}}{
  2\omega_{\bar{k}}(t)}+ \frac{e^{-2 \Theta^2(t)}}{4 i}
   \left[\frac{\tilde{\Omega}_{+,\kappa}}{
     2\omega_{\kappa}(t)}e^{2 i \theta^-_{\bar{k}}(t)}-
     \frac{\tilde{\Omega}_{+,\kappa^*}}{
     2\omega_{\kappa^*}(t)}e^{-2 i \theta^-_{\bar{k}}(t)}
  \right]\right\} +Q(m,t),
\end{eqnarray}
where $Q(m,t)=\int_{|{\bf k}|>m_f} G^S_{\bf k}(t)- I_0(m^2)$, of
which we present a rough estimation in appendix C. We do not need
the full expression of $Q(m,t)$, rather it is enough to know that
it behaves as $Q(m,t) \propto -m_f^2 \sqrt{x/x_f}$ for large $x$.

Other integrals in Eq.~(\ref{Hfin}) are given by
\begin{eqnarray} \label{integrals}
&&\int_{|{\bf k}|<m_f}\frac{G^S_{\bf k}(t)}{\omega_{\bf k}^4(t)}
\simeq
  \bar{l}\left\{ \frac{\tilde{\Omega}_{+,\bar{k}}}{2
  \omega_{\bar{k}}^5(t)}
  +\frac{e^{-2 \Theta^2(t)}}{2 i}
   \left[\frac{\tilde{\Omega}_{+,\kappa}}{
     2\omega_{\kappa}^5(t)}e^{2 i \theta^-_{\bar{k}}(t)}-
     \frac{\tilde{\Omega}_{+,\kappa^*}}{
     2\omega_{\kappa^*}^5(t)}e^{-2 i \theta^-_{\bar{k}}(t)}
  \right]\right\} , \\
&&\int_{|{\bf k}|<m_f} \frac{g_{\bf k}^- R_{\bf k}^-}{\omega_{\bf
    k}^2(t)}
    \sin \left[2\theta^-_{\bf k}(t)-\psi_{\bf
    k}^-\right]
 \simeq -\bar{l} e^{-2 \Theta^2(t)}
    \left[\frac{\tilde{\Omega}_{+,\kappa}}{
     2\omega_{\kappa}^2(t)}e^{2 i \theta^-_{\bar{k}}(t)}+
     \frac{\tilde{\Omega}_{+,\kappa^*}}{
     2\omega_{\kappa^*}^2(t)}e^{-2 i \theta^-_{\bar{k}}(t)}
  \right]  ,\\
&&\frac{1}{8} \int_{+,\bf k}\left[1-g_{\bf k}^{\pm 2}(1- R_{\bf
k}^{\pm 2} )\right] (G_{\bf k}^S)^{-1} \simeq  0 ,
\end{eqnarray}
where in the second integral we use $\psi^-_{\bf k}\simeq \pi/2$,
$g^-_{\bf k} \simeq \tilde{\Omega}_{+,\bf k}e^{2
\tilde{\theta}_{\bf k}(t_k)}$, and $R^-_{\bf k} \simeq 1$. We
ignore the integrals $\displaystyle \int_{|{\bf
k}|>m_f}\frac{G^S_{\bf k}(t)}{\omega_{\bf k}^4(t)}$ and
$\displaystyle \int_{|{\bf k}|>m_f} \frac{g_{\bf k}^- R_{\bf
k}^-}{\omega_{\bf k}^2(t)}
    \sin \left[2\theta^-_{\bf k}(t)-\psi_{\bf k}^-\right]$
compared to its small momentum term in Eq.~(\ref{integrals}).

Therefore, the $H_f$ becomes
\begin{eqnarray} \label{H_f:large t}
H_f&\simeq & \frac{1}{2}\dot{q}^2 + \frac{\bar{l}
\tilde{\Omega}_{+,\bar{k}}}{4\omega_{\bar{k}}} \left[ m^2 +
\frac{(\dot{m^2})^2}{16 \omega_{\bar{k}}^4} \right]
+\frac{1}{2}m^2 Q(m,t) \\
&+& \frac{\bar{l}e^{-2 \Theta(t)^2}}{8}\left\{
  \frac{(\dot{m^2})^2}{8i}\left[\frac{\tilde{\Omega}_{+,\kappa}}{
     2\omega_{\kappa}^5(t)}e^{2 i \theta^-_{\bar{k}}(t)}-
     \frac{\tilde{\Omega}_{+,\kappa^*}}{
     2\omega_{\kappa^*}^5(t)}e^{-2 i \theta^-_{\bar{k}}(t)}
 \right]-\dot{m^2}\left[\frac{\tilde{\Omega}_{+,\kappa}}{
     2\omega_{\kappa}^2(t)}e^{2 i \theta^-_{\bar{k}}(t)}+
     \frac{\tilde{\Omega}_{+,\kappa^*}}{
     2\omega_{\kappa^*}^2(t)}e^{-2 i \theta^-_{\bar{k}}(t)}
  \right] \right. \nonumber \\
 & -& \left. im^2\left[\frac{\tilde{\Omega}_{+,\kappa}}{
     2\omega_{\kappa}(t)}e^{2 i \theta^-_{\bar{k}}(t)}-
     \frac{\tilde{\Omega}_{+,\kappa^*}}{
     2\omega_{\kappa^*}(t)}e^{-2 i \theta^-_{\bar{k}}(t)}
  \right]
 \right\} \nonumber.
\end{eqnarray}
From Eq.~(\ref{intermediate}) we know that the last integral is
finite and small compared to the exponentiated terms. Similarly,
the integral $\int_{|{\bf k}|>m_f}\frac{G_{\bf k}}{ \omega_{\bf
k}^4} $ can also be ignored compared to other terms.

We approximate $V_\tau$ also by its value around $k=m_f$. We
should also calculate $\displaystyle \frac{1}{2}\int_{+,\bf k}
g_{\bf k}^- \omega_{\bf k}$ in Eq.~(\ref{H:ren}), which gives
\begin{eqnarray} \label{Vtau}
V_\tau = \frac{\bar{l}}{2}\tilde{\Omega}_{+,
\bar{k}}\omega_{\bar{k}}(t)+
\frac{m_f^4}{32\pi^2}\left[\sqrt{1+\frac{x}{x_f}}
 \left(2+\frac{x}{x_f}\right)-\frac{x^2}{x_f^2}\ln
 \left(\sqrt{\frac{x_f}{x}}+\sqrt{1+\frac{x_f}{x}}\right) \right]+
 \frac{m_f^4g'(x_i,x_f)}{32 \pi^2}  \sqrt{\frac{x}{x_f}+1} ,
\end{eqnarray}
where $g'$ is a positive constant and the integrals needed in this
equation is calculated in Appendix C.

One can sum everything to obtain $\displaystyle H_{eff}=D'+H_{f}
+V_{\tau}+ \frac{m_R^4}{64\pi^2} {\cal V}_{vac}$. Before doing
this summation, let us investigate where does the main
contributions come from. Since $\omega_{\bar{k}}\simeq m(t)$, it
is enough to observe the $m(t)$ dependence of each terms. The main
contribution comes from the unstable modes which have $\bar{l}$
factor if the value of $x$ is not very large. For larger value of
$x$ the $-x^2 \ln |x|$ term in $V_{vac}$ starts to compete with
$\bar{l}$ term and then dominates the potential for larger $x$.
Therefore, to know the behavior of the effective Hamiltonian, we
need to keep these terms only:
\begin{eqnarray} \label{H:effFinal}
H_{eff}-D'&\simeq & K+ \frac{\bar{l}\tilde{\Omega}_{+,\bar{k}}}{2}
\left[\omega_{\bar{k}}(t) + \frac{m^2(t)}{2
\omega_{\bar{k}}(t)}\right] \\
&+&
   \frac{\bar{l}m^2(t)e^{-2\Theta^2(t)}}{8i}\left[\frac{\tilde{\Omega}_{+,\kappa}}{
     2\omega_{\kappa}(t)}e^{2 i \theta^-_{\bar{k}}(t)}-
     \frac{\tilde{\Omega}_{+,\kappa^*}}{
     2\omega_{\kappa^*}(t)}e^{-2 i \theta^-_{\bar{k}}(t)}
  \right]-
\frac{m_R^4}{64\pi^2} x^2\ln x , \nonumber
\end{eqnarray}
where $K$ represents the sum of all kinetic terms and
$\omega_{\bar{k}}/m_R= \sqrt{x+ \bar{x}} $ with
$\bar{x}=\bar{k}^2/m_R^2$. Therefore, the effective potential for
very large $t$ and $x>0$ becomes
\begin{eqnarray} \label{V:inf}
{\cal V}=\frac{V_{eff}-D'}{m_R^4/(64\pi^2)}&\simeq &
\frac{2v}{3}\left(\sqrt{x+ \bar{x}}+ \frac{x}{2 \sqrt{x+
\bar{x}}}\right)- x^2\ln x ,
\end{eqnarray}
where $\displaystyle v=\frac{24 \pi^2\bar{l}(x_i+x_f)}{m_R^3
\sqrt{x_i x_f}}$, and $\bar{x} \ll 1$ for large $\bar{t}$. Note
that a natural stabilization of the effective potential occurs for
large $t>{\cal T}$ due to the factor $e^{-2\Theta^2(t)}$.

Until now, we did not calculate the effective action for negative
$x$ which is impossible until we treat the IR divergences
properly, but the effective potential  can still be calculable
which exponentially increases in time as can be seen in Sec. III.
The main contribution of the unstable modes comes from the
integral $\int_{{\bf k}^2<-m^2}\tilde{\omega}_{\bf k}^2G^U_{\bf
k}$ which becomes
\begin{eqnarray} \label{-m}
\int_{{\bf k}^2<-m^2}\tilde{\omega}_{\bf k}^2G^U_{\bf k} &\simeq&
\frac{1}{8 \pi^2} \int_0^{\bar{m}} dk \tilde{\omega}_{\bf k}(t)
\tilde{\Omega}_{+, \bf k} k^2e^{2 \tilde{\theta}_{\bf k}(t)} \\
&\simeq& \frac{\tilde{\omega}_{{\bf k}_t}\tilde{\Omega}_{+,{\bf
k}_t}}{4} \left(\frac{\pi {\bf k}_t^2}{2 \pi}\right)^{3/2} e^{2
\tilde{\theta}_{{\bf k}_t}(t)} , \nonumber
\end{eqnarray}
where $k_t$ is determined by $\displaystyle {\bf k}_t^2=\int_0^t
dt'\tilde{\omega}_{{\bf k}_t}^{-1}(t')$. Due to the exponential
factor in Eq.~(\ref{-m}), the potential~(\ref{-m}) for large $t$
is very sharply inclined to the vertical axis for negative $x$ so
that $x$ increases. In summary, for a given time $t$, the
effective potential for $x$ has minimum at $x=0$ and increases in
both directions. For large $x\sim v^{2/3}$, the $x^2 \ln x$ term
starts to compete with $v\sqrt{x}$. For larger $x> v^{2/3}$, the
potential starts to decrease. The relation between
$\bar{q}^2-\bar{q}_R^2$ and $x$ is still determined by
Eq.~(\ref{x:Phi}), whose solution is obtained by graphical method
shown in Fig. 3 where the form of the potential is different from
that in Fig. 1.
\begin{figure}[htbp]
\begin{center}
\includegraphics[width=.6\linewidth,origin=tl]{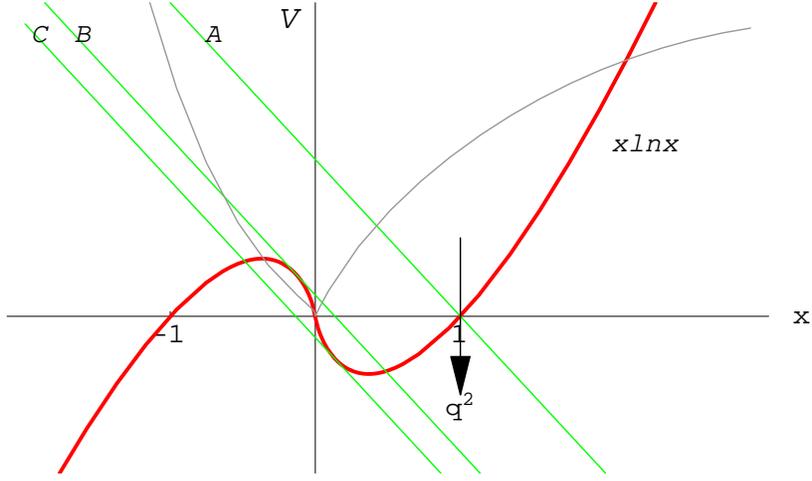}\hfill%
\end{center}
\caption{Graphical solution for Eq.~(\ref{x:Phi}). The thick curve
represents the function $x \ln |x|$ and the thin curve represents
the effective potential ${\cal V}/v$ (we set $v=50$ and
$\bar{x}=1/4$ and normalize it so that it goes to zero at $x=0$.)
at a given time, and the straight lines A, B, C represent the left
hand side of (\ref{x:Phi}) with $\Phi^2-\Phi_R^2=0,~
\kappa-1-e^{-\kappa},~ \kappa -1 + e^{-\kappa}$, respectively, for
$\kappa >1$. The arrow indicates how the lines move as
$\Phi^2-\Phi_R^2$ increases. The potential has minimum at $x=0$.}
\label{qgt:fig}
\end{figure}
With these we schematically plot the potential in $\bar{q}$ in
Fig. 4,
\begin{figure}[htbp]
\begin{center}
\includegraphics[width=.5\linewidth,origin=tl]{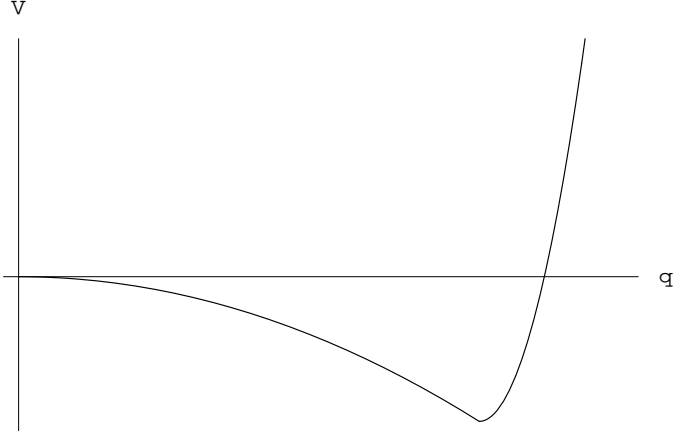}\hfill%
\end{center}
\caption{Typical form of the effective potential ${\cal V}$ at a
given time $t$. The horizontal axis is $\bar{q}$. The global
minimum is at $\bar{q}\neq 0$ and $q=0$ is a local maximum.}
\label{pot2:fig}
\end{figure}
which shows the presence of the symmetry breaking clearly. A few
comments are in order. First, the potential is plotted as a
function of $\bar{q}$. Since the minimum value of $\bar{q}$
corresponds to $q=0$, the form of the potential can roughly be
interpreted as a potential for $q$ if $m^2>0$. On the other hand,
if $m^2<0$, the relation between $\bar{q}$ and $q$ changes in
time. Therefore, we should interpret the potential only at a given
time. Second, the restriction $\kappa >0$ in the
potential~(\ref{VD:nor2}) disappears in the present case. Even for
negative $\kappa$, the solution at $x=1$ corresponds to the
smaller potential value of the two positive solution
of~(\ref{x:Phi}) for large enough $v$.

All of the discussions in this section are the zero temperature
results. The inclusion of non-zero initial temperature does not
leads to a critical complication in the analysis. The only
difference is that all the integrals in the calculations should
contains the $\displaystyle \coth \frac{\beta \omega_{i,\bf
k}}{2}$ factor. This results in the change of $\displaystyle
\bar{l} \longrightarrow \bar{l}\coth \frac{\beta
\omega_{i\bar{k}}}{2}$ for the unstable modes and the ${\cal V}$
in Eq.~(\ref{VD:nor2}) becomes temperature dependent. However,
these cannot alter the global behavior of the effective potential.

\section{Equal time correlation function}
Until now we have calculated the effective action with the
precarious renormalization condition. However, the correlation
function,
\begin{eqnarray} \label{Gr}
G({\bf x},{\bf y};t)&=& \int\frac{d^3 {\bf k}}{(2\pi)^3} e^{i {\rm
k}\cdot ({\bf x-y})} G_{\bf k}(t) \\
&=& \frac{1}{2 \pi^2 r}\int k dk~ \sin kr ~G_{\bf k}(t)  \nonumber
\\
&\simeq& \frac{1}{2 \pi^2 r}\left[\int_0^{\bar{m}}k dk~ \sin kr
 ~G^U_{\bf k}(t)+\int_{\bar{m}}^{m_f} k dk~ \sin kr ~G^S_{\bf
  k}(t)+ \int_{m_f}^{\infty}k dk~ \sin kr ~G^S_{\bf
  k}(t)\right], \nonumber
\end{eqnarray}
where $r=|{\bf x-y}|$, is renormalization scheme independent in
the sense that the renormalization is related only to the stable
modes and the dominant contribution to the correlation function
comes from the exponential growth of the unstable modes. This
observation enables one to ignore the last integral in
Eq.~(\ref{Gr}) which is related to the two point correlation
function of stable system, which we ignore in this section
compared to the first two integrals which contain exponentially
increasing factors for large time.

Let us consider the first integral of Eq.~(113) where the dominant
part of the integrand is proportional to $k^2e^{2
\tilde{\theta}_{\bf k}(t)}$, which is peaked around $\bar{k}$
where $\bar{k}^{-2}=\int_0^t dt'\tilde{\omega}^{-1}_{\bar{k}}( t')
$. Then, the correlation function~(\ref{Gr}) can be approximated
by the contribution of the unstable mode $G^U$,
\begin{eqnarray} \label{int:ap}
G^U({\bf x,y};t)&=&\frac{1}{2 \pi^2}\int_0^{\bar{m}}k^2 dk~
\frac{\sin kr}{kr}
 ~G^U_{\bf k}(t)\simeq\frac{1}{8\pi^2}
 \int_0^{\bar{m}} dk~ \frac{\tilde{\Omega}_{+,\bf
 k}}{\tilde{\omega}_{\bf k}} k^2 e^{2 \tilde{\theta}_{\bf k}(t)}
 \frac{\sin kr}{kr} \\
&\simeq&\left(\frac{\bar{k}^2}{2\pi}\right)^{3/2} \exp\left[2
 \tilde{\theta}_{\bar{k}}(t)-\frac{\bar{k}^2r^2}{8}\right]
   \left[\frac{\tilde{\Omega}_{+,\kappa}}{\kappa
    \tilde{\omega}_{\kappa}} \frac{e^{i \bar{k} r}}{2ir}-
   \frac{\tilde{\Omega}_{+,\kappa^*}}{\kappa^*
    \tilde{\omega}_{\kappa^*}} \frac{e^{-i \bar{k} r}}{2ir}
    \right]
   , \nonumber
\end{eqnarray}
where $\displaystyle \kappa = \bar{k}+ i \frac{r}{2 \gamma}$. For
large $t$, $\bar{k}/\bar{m}$ becomes very small, and the
coefficient of $k^2e^{2 \tilde{\theta}_{\bf k}(t)}$ should be
smooth for small $k$. Another consequence of the small $\bar{k}$
is that the Gaussian integral well approximates $G^U$ for most
range of $0<t<{\cal T}$, until $\bar{m}> \bar{k}$. This is why we
approximate $k^2e^{2 \tilde{\theta}_{\bf k}(t)}$ as a Gaussian
function (the steepest descent method) factoring out the smooth
function $\displaystyle \frac{\sin kr}{kr}$ in the integrand.

Let us consider the $r \rightarrow 0$ limit of this function,
\begin{eqnarray} \label{r-0}
\lim_{r\rightarrow 0}G^U({\bf x,y};t)&=&\left(\frac{\bar{k}^2}{2
  \pi}\right)^{3/2} \frac{e^{2
 \tilde{\theta}_{\bar{k}}(t)}}{4}
   \left[\frac{\tilde{\Omega}_{+,\bar{k}}}{
    \tilde{\omega}_{\bar{k}}}+ \frac{1}{2}\lim_{r\rightarrow 0}
     \Im \frac{\tilde{\Omega}_{+,\kappa}}{
    r \kappa\tilde{\omega}_{\kappa}} \right] \\
&=& \frac{7}{8}\left(\frac{\bar{k}^2}{2
  \pi}\right)^{3/2}\frac{e^{2
 \tilde{\theta}_{\bar{k}}(t)}}{4}
   \frac{\tilde{\Omega}_{+,\bar{k}}}{
    \tilde{\omega}_{\bar{k}}} =\frac{7}{8} G^U({\bf
    x,x};t)=\frac{7}{8}G(0;t)
      \nonumber ,
\end{eqnarray}
where in the last equality we use $\displaystyle
\lim_{r\rightarrow 0}
     \Im \frac{\tilde{\Omega}_{+,\kappa}}{
    r \kappa\tilde{\omega}_{\kappa}} = \frac{1}{2\gamma}
    \frac{\partial}{\partial \bar{k}} \left(\frac{
     \tilde{\Omega}_{+,\bar{k}}}{ \bar{k}
    \tilde{\omega}_{\bar{k}}}\right) \simeq -\frac{
     \tilde{\Omega}_{+,\bar{k}}}{ \bar{k}^2
    \tilde{\omega}_{\bar{k}}} $.
At $r=0$, this should correspond to $G^U({\bf x,x};t)=\int_{|{\bf
k}|<\bar{m}} G_{\bf k}^U(t)$. Therefore in general we write
\begin{eqnarray} \label{G:r}
G({\bf x,y};0<t<{\cal T})&\simeq& \frac{7}{16} G^U(0;t)
\frac{\bar{\omega}_{\bar{k}}(t)}{\tilde{\Omega}_{+,\bar{k}}}
\exp\left(-\frac{\bar{k}^2r^2}{8}\right)
   \left[\frac{\tilde{\Omega}_{+,\kappa}}{r\kappa
    \tilde{\omega}_{\kappa}(t)} e^{i \bar{k} r}-
   \frac{\tilde{\Omega}_{+,\kappa^*}}{r\kappa^*
    \tilde{\omega}_{\kappa^*}(t)} e^{-i \bar{k} r} \right] ,
\end{eqnarray}
where most of time-dependencies are given by $G^U(0;t)$ in
Eq.~(\ref{r-0}). Note that $1/\bar{k}^2 \simeq 2\int_0^t dt'
\tilde{\omega}_{\bar{k}}^{-1}(t')$ increases as time. The exponent
$\bar{k}^2r^2/8$ determines the correlation length,
\begin{eqnarray} \label{xi}
\xi_D(t)= 2\sqrt{2} \left[\int_0^t dt'
\tilde{\omega}_{\bar{k}}^{-1}(t')\right]^{1/2}\simeq
2\sqrt{2}\left[\int_0^t dt' \tilde{m}^{-1}(t')\right]^{1/2} ,
\end{eqnarray}
which gives the Cahn-Allen relation for constant $\bar{m}^2$, time
lagged deformed relation $\displaystyle
4\left[\frac{1}{\alpha}\left(m_f-\sqrt{m_f^2-\alpha t}
\right)\right]^{1/2}$ for linear mass-squared, $\bar{m}^2(t)=m_f^2
-\alpha t$, which was discovered in Ref.~\cite{spkim}, and another
deformed relation $\displaystyle 2 \left[\frac{2}{\alpha}
\tan^{-1}\frac{\alpha t}{\sqrt{m_f^2-\alpha^2 t^2}}\right]^{1/2}$
for quadratic mass-squared $\bar{m}^2(t)=m_f^2-\alpha^2 t^2$. Note
that the last two deformed relations have the maxima of the
correlation length given by $\displaystyle
\sqrt{\frac{8m_f}{\alpha}}$ and $\displaystyle \sqrt{\frac{2
\pi}{\alpha}}$, respectively. This is because the instability ends
to increase at the time where $\bar{m}^2(t)=0$. An interesting
feature of the last relation is that the maximum correlation
length does not depend on the initial mass-squared $m_f^2$,
rather, it depends only on the acceleration of the change of the
mass-squared. Since we do not know the exact time-dependence of
$m^2(t)$ due to the limitation of WKB method mentioned in Sec. V,
we cannot predict the exact form of the correlation length.

Next, let us consider the second integral of Eq.~(\ref{Gr}). The
dominant part of the integral is proportional to $\bar{k}^2 e^{2
\tilde{\theta}_{k}(t_k)}$, which is peaked at $\displaystyle
\bar{k}^{2}=\left(\frac{m_f}{\alpha \bar{t}}\right)^{1/2}$. The
approximation of this integral is illustrated in Appendix B. Since
the integral is maximized at $\bar{k} \ll m_f$, we can ignore this
term until $\bar{m}(t)\sim \bar{k} \sim 0$. Therefore, during the
process of instability growing, $0< t< {\cal T}$, we can ignore
the second integral. In this sense, this integral becomes
important only after $t> {\cal T}$ since we have no $G^U$ term in
this region. The dominant contribution to $G({\bf x,y};t)$ comes
from the second integral for $t>{\cal T}$ and it gives
\begin{eqnarray} \label{i2}
\int_{0}^{m_f} k dk~ \sin kr ~G^S_{\bf
  k}(t)&=&\frac{1}{2\pi^2}\int_0^{m_f}dk k^2e^{2
  \tilde{\theta}_{\bf k}(t_k)}\frac{\tilde{\Omega}_{+,\bf k}}{
    2 \omega_{\bf k}}  \frac{\sin kr}{kr} \left[1+ \sin 2
    \theta^-_{\bf k}(t) \right] \\
&=& \left(\frac{m_f}{2 \pi \alpha
  \bar{t}}\right)^{3/2} \exp\left[2\tilde{\theta}_{\bar{k}}(\bar{t})
    - \frac{\bar{k}^2r^2}{8}\right] \left[
  g\left(\bar{k}+ i\frac{\bar{k}^3r}{4}\right) \frac{e^{i
  \bar{k}r}}{2ir}
  -g\left(\bar{k}- i\frac{\bar{k}^3r}{4}\right) \frac{e^{-i
  \bar{k}r}}{2ir}
 \right] \nonumber \\
&-& \left(\frac{m_f}{2 \pi \alpha
  \bar{t}}\right)^{3/2}e^{2\tilde{\theta}_{\bar{k}}(\bar{t})}
  \left\{ e^{-2 \Theta^2(t,r)}
   \Re \left( g\left[\bar{k}+i\bar{k}\Theta(t,r)\right]
  \frac{e^{ i[2\theta^-_{\bar{k}}(t) + \bar{k}r]}}{2r}
   \right) \right.
     \nonumber\\
&-&\left. e^{-2 \Theta^2(t,-r)}
   \Re \left( g\left[\bar{k}+i\bar{k}\Theta(t,-r)\right]
  \frac{e^{ i[2\theta^-_{\bar{k}}(t) - \bar{k}r]}}{2r} \right)
  \right\} , \nonumber
\end{eqnarray}
where $\displaystyle g(k)=\frac{\tilde{\Omega}_{+,\bf k}}{2k
\omega_{\bf k}}$ and $\Phi(t,r)$ is given in Eq.~(\ref{Theta}).
Because of the exponential decaying factor, most terms of $G({\bf
x,y};t )$ decrease to zero for large $r$ and $t$. However, a
non-trivial exception exists which signals the end of the spinodal
line along the line $\Theta(t,-r)=0$ since, on this line the
exponential decaying factor $e^{-2 \Theta^2(t,-r)}$ of the last
line in Eq.~(\ref{i2}) disappears, which gives the non-trivial
long time behavior of the correlation function after symmetry
breaking,
\begin{eqnarray} \label{long time}
G({\bf x,y};t) &\sim& \left(\frac{m_f}{2 \pi \alpha
  \bar{t}}\right)^{3/2} \exp\left[2\tilde{\theta}_{\bar{k}}(\bar{t})
    - \frac{\bar{k}^2r^2}{8}\right] \left[
  g\left(\bar{k}+ i\frac{\bar{k}^3r}{4}\right) \frac{e^{i
  \bar{k}r}}{2ir}
  -g\left(\bar{k}- i\frac{\bar{k}^3r}{4}\right) \frac{e^{-i
  \bar{k}r}}{2ir}
 \right] \\
 &+& \left(\frac{m_f}{2 \pi \alpha
  \bar{t}}\right)^{3/2} \frac{\tilde{\Omega}_{+,\bar{k}}}{2
  \omega_{\bar{ k}}}
 e^{2\tilde{\theta}_{\bar{k}}(\bar{t})}
  e^{-2 \Theta^2(t,-r)} \frac{\cos\left[2
  \theta^-_{\bar{k}}(t)-\bar{k}r\right]}{\bar{k}r} .  \nonumber
\end{eqnarray}
The first term is the usually expected time-independent
correlation, which corresponds to a domain formation and growth.
The second term of Eq.~(\ref{long time}) is a new time-dependent
correlation, which propagates even after the end of the phase
transition, and travels along the line $\displaystyle
\bar{k}r=\frac{2m_f}{\alpha \bar{t}} \int_{\bar{t}}^t
\frac{dt'}{\omega_{\bar{k}}(t')}$. This term may be interpreted as
a propagating domain.

\section{Summary and discussions}
In this paper we have elaborated the extension of the double
Gaussian type wave-functional approximation to study the
non-equilibrium quantum dynamics of self-interacting scalar field
system in (3+1) dimensions. The time-dependent second order phase
transition is a classic example of this type of problems. These
systems are characterized by time-dependent coupling parameters
and their true nonequilibrium evolution deviates significantly
from the equilibrium one when their coupling parameters differ
greatly from their initial values. In this case the systems evolve
completely out of equilibrium. To understand the dynamical aspects
of these processes there have been developed many different
methods such as the closed-time path integral method, sometimes in
conjunction with the large $N$ expansion, mean-field,
Hartree-Fock, and the Liouville-von Neumann methods. The present
paper uses the functional-Schr\"{o}dinger method and WKB
approximation in connection with the Liouville-von Neumann
approach to obtain the effective action of explicitly
time-dependent system undergoing phase transition.

By applying this method to the scalar $\lambda \phi^4$ theory
undergoing second order phase transition, we have found the
effective equations of motion and effective action of the system
in terms of the correlations of the field. This equation of motion
was used to obtain the effective Hamiltonian of the system written
in terms of the mode solutions, which is the starting point of the
present method. To have better approximation of the theory, we
first apply the method to the case of constant mass-squared
including the tachyonic modes. To do this, we renormalized the
effective action by employing the standard precarious
renormalization scheme and show that the effective potential has
no symmetry breaking even if one includes the tachyonic modes. The
inclusion of the tachyonic modes only results in the presence of a
new metastable state at the zero mass region. Since what we would
like to describe is the system with time-dependent mass-squared
which evolves in time self-consistently, we added the exact time
dependent mode solutions under a prescribed change of the
mass-squared. By extracting the limiting behaviors of this mode
solutions we get the WKB approximated solutions of the general
time-dependent self-interacting scalar field system with
instantaneous quenching at $t=0$. The full renormalization of the
equation of motion and the effective action was performed to
obtain the finite dynamical evolution of the zero-mode and the
mass-squared. By analyzing the time-independent part of the
effective potential we show that the vacuum structure is changed
from that of the system without quenching. It is shown that the
effective potential indicates the existence of the second order
phase transition, which have a new vacuum at the non-zero position
of the field expectation value and the position of zero
expectation value corresponds to the unstable equilibrium.

Due to the limitation of the WKB approximation, we cannot predict
the time-dependence of mass-squared if it is negative. On the
other hand, the effective potential and the spatial correlation
function are calculable. For these calculations, we develop a
large instability approximation which takes into account the
exponential increase of the unstable modes dominating the
Hamiltonian after renormalization. We computed the spatial
correlation function with this approximation. Before the phase
transition ends, the correlation is dominated by the unstable
modes with very low frequency determined by the formula $
\bar{k}=\left[\int_0^t dt'\tilde{m}^{-1}( t')\right]^{-1/2} $,
which decreases as time. The inverse of $\bar{k}$ corresponds to
the correlation length of the system. We show that the classical
Cahn-Allen relation holds only when the mass-squared is a negative
constant and we have shown that the correlation length depends in
general on the time-dependence of the mass-squared. An interesting
addition in this paper is that the maximum correlation length
exists depending on the time-dependence of the mass-squared.
Especially, if the mass-squared increases quadratically in time,
the maximal correlation length is independent of the initial
mass-squared. Once the phase transition process ends, the spatial
correlation is dominated by the stable modes with low frequency.
Besides the usual static correlation corresponding to the
formation and growth of domains, it is shown that there exists a
travelling correlations. This correlation starts to expand
spherically when the phase transition ends with the travelling
velocity $\displaystyle v\sim 2 \sqrt{\frac{m_f}{\alpha
\bar{t}}}\frac{1}{\omega_{\bar{k}}(t)}$. The spatial size of this
correlation is $\displaystyle \frac{2 \sqrt{2}}{\bar{k}}$.
Therefore, this correlation is separated from the static one in
time $\displaystyle \frac{\omega_{\bar{k}}}{m_f} \alpha \bar{t}$
where $\bar{t} \sim {\cal T}$. The presence of this correlation
can be tested on experiment.

\begin{acknowledgments}
This work was supported in part by Korea Research Foundation under
Project number KRF-2003-005-C00010 (H.-C.K. and J.H.Y.).
\end{acknowledgments}
\vspace{3cm}

\begin{appendix}

\section{long time approximation}
The function $f(y)$ of Eq.~(36) is given by
\begin{eqnarray}
 f(y)&=& 2 \bar{m}t (1-y)^{1/2}+ \ln y - \frac{1}{2} \ln (1-y)+
 \ln \bar{m} ,  ~~ 0 \leq y \leq 1.
\end{eqnarray}
Its first and second derivatives become
\begin{eqnarray} \label{f'}
f'(y) &=& \frac{1}{2 y (1-y)} \left[2-y - 2\bar{m} t y
\sqrt{1-y}\right], \\
f''(y) &=& -\frac{\bar{m}t}{2(1-y)^{3/2}}- \frac{1}{y^2}+
\frac{1}{2(1-y)^2}. \label{f''}
\end{eqnarray}
The function $g(y)=2-y-2\bar{m}t y\sqrt{1-y}$ in the parenthesis
of $f'$ has 2 roots in the defined range of $y$ since $g(0)=2$,
$g(1)=1$, and $g(1/2)<0$ for large $\bar{m}t$. Since $g(y)$
decreases at $y=0$, smaller solution to this corresponds to the
maximum value of $f(y)$. In the large $t$ approximation, the
smaller root is $\displaystyle \bar{y}=\frac{1}{\bar{m} t}$. Then,
$f(y)$ can be series expanded as
\begin{eqnarray} \label{f:1}
f(y)\simeq f\left(\frac{1}{\bar{m} t}\right) +
\frac{1}{2}f''\left(\frac{1}{\bar{m}t}\right)
\left(y-\frac{1}{\bar{m} t}\right)^2+ \cdots =
f\left(\frac{1}{\bar{m} t}\right) - \frac{(\bar{m}t)^2}{2}
\left(y-\frac{1}{\bar{m} t}\right)^2+ \cdots.
\end{eqnarray}
Then, the integral for slowly varying function $g(k)$ becomes
\begin{eqnarray} \label{fn}
\int dk g(k) e^{f(k^2/\bar{m}^2)} &\simeq &
g(\bar{k})e^{f(\bar{k}^2/\bar{m}^2)} \int dk
\exp\left[-\frac{2t}{\bar{m}}
\left(k-\sqrt{\frac{\bar{m}}{t}}\right)^2\right] \\
&\simeq &\sqrt{\frac{\pi \bar{m}}{2t^3}}~
g(\bar{k})e^{2\bar{m}t-1} . \nonumber
\end{eqnarray}

\section{large instability approximation}
The first step of the large instability approximation is to
approximate $f(y)$ of Eq.~(99) up to quadratic part around the
maximum point of it:
\begin{eqnarray} \label{f:y}
f(y)=2 m_f \int_0^{t_k} \sqrt{\nu(t')-y} dt' + \ln y + 2 \ln m_f =
f(\bar{y})- \frac{\gamma}{2} (y-\bar{y})^2+ \cdots,
\end{eqnarray}
where  $\nu(t)= \bar{m}^2(t)/m_f^2$, $(\bar{y},
\bar{t}=t_{\bar{k}})$ is the point where $f(y)$ is maximum
determined by
\begin{eqnarray} \label{peak2}
 \frac{1}{\bar{y}}=m_f\int_0^{\bar{t}}
 \frac{dt}{ \sqrt{\nu(t)-\bar{y}}} ,
\end{eqnarray}
and $\gamma$ is the negative of the curvature of $f$ at that
point:
\begin{eqnarray} \label{gamma1}
\gamma =\left.-\partial_y^2f(y)\right|_{y=\bar{y}}=
\frac{1}{\bar{y}^2}+ \frac{m_f}{2}\lim_{t \rightarrow
t_k}\left[\int_0^t \frac{dt'}{(\nu(t')-y)^{3/2}}-
\frac{2}{\dot{\nu}(t)}\frac{1}{\sqrt{\nu(t)-y}}
\right]_{t_k=\bar{t}} ,
\end{eqnarray}
which is $O(m_f^2 \bar{t}^2)$. Let us approximately estimate the
size of $\bar{y}$. Let the rate of change of the mass squared
$\dot{m^2}(t=\tau)=0$. We can safely assume that the acceleration
$\displaystyle \frac{d^2m^2(t)}{dt^2}$ is non-negative during
$\tau<t< {\cal T}$, which can be easily understood from the
potential in Fig. 1, where all unstable modes tend to increase the
mass-squared if it is negative. Therefore, with
$\nu(t)=\bar{m}^2(t)/m_f^2$, $\nu(t)-y$ satisfies $(1-y)(1-t/t_k)
\leq \nu(t)-y \leq 1-y$ since its initial value is $\nu(t=\tau)=1$
and $\nu(t_k)-y=0$. With this inequality we get
\begin{eqnarray} \label{exponents}
 &&\frac{2t_k}{3}\sqrt{1-y} \leq \int_0^{t_k} dt
 \sqrt{\nu(t)-y}  \leq t_k\sqrt{1-y} , \\
&& \frac{t_k}{
 \sqrt{1-y}} \leq \int_0^{t_k}
 \frac{dt}{ \sqrt{\nu(t)-y}} \leq \int_0^{t_k}
 \frac{dt}{ \sqrt{(1-y)(1-t/t_k)}}=\frac{2 t_k}{
   \sqrt{1-y}} ,
\end{eqnarray}
which leads to,
\begin{eqnarray} \label{tkk}
m_f^2\bar{y}=\bar{k}^2 = \frac{m_f}{\alpha\bar{t}} ,
\end{eqnarray}
with $1\leq \alpha \leq 2$. With this result, $\gamma \simeq
\alpha^2 m_f^2 \bar{t}^2$.

Since this $\gamma$ is large, we can approximate
\begin{eqnarray} \label{g:approx}
\int dk g(k)k^2 e^{2 \tilde{\theta}_{\bf k}(t_k)} &\simeq &
g(\bar{k}) \bar{k}^2 e^{2 \tilde{\theta}_{\bar{k}}(\bar{t})} \int
dk \exp\left[-\frac{2 \alpha
\bar{t}}{m_f}\left(k-\sqrt{\frac{m_f}{\alpha \bar{t}}}\right)^2
\right]  \\
&\simeq & \sqrt{\frac{\pi}{2}}\left(\frac{m_f}{\alpha
\bar{t}}\right)^{3/2} g(\bar{k})e^{2
\tilde{\theta}_{\bar{k}}(\bar{t})} . \nonumber
\end{eqnarray}

Series expanding $\int_{t_k}^t\omega_{\bf k}(t') dt'$ to first
order in $(k-\bar{k})$ we have,
\begin{eqnarray} \label{omega}
\int_{t_k}^t\omega_{\bf k}(t') dt'&=& \int_{t_k}^{\bar{t}}
  \omega_{\bf k}(t') dt'+\int_{\bar{t}}^t\left[\omega_{\bf k}(t')-
  \omega_{\bar{k}}(t')\right]dt'+
  \int_{\bar{t}}^t\omega_{\bar{k}}(t')dt' \\
&\simeq &
\left(\bar{k}\int_{\bar{t}}^t\frac{dt'}{\omega_{\bar{k}}(t')}
\right)(k-\bar{k}) + \int_{\bar{t}}^t \omega_{\bar{k}}(t') dt' ,
\nonumber
\end{eqnarray}
where we ignore $\int_{t_k}^{\bar{t}} \omega_{\bf k}(t') dt'$
since it is $O(\bar{k})$ for all time and we use $\displaystyle
\bar{\omega}_{\bf k} \simeq \omega_{\bar{k}}+
\frac{\bar{k}}{\omega_{\bar{k}}}(k-\bar{k}) + \cdots$ in the
second equality. Using this equation we get
\begin{eqnarray} \label{g:wr}
\frac{2\alpha \bar{t}}{m_f}(k-\bar{k})^2\pm 2i \theta^-_{\bf k}(t)
\pm i k r
  &\simeq&
 \frac{2\alpha \bar{t}}{m_f}
  \left[k-\bar{k}\pm i\bar{k}\Theta(t,r)\right]^2 +2\Theta^2(t,r)
 \pm i\left[\bar{k}r+ 2\theta^-_{\bar{k}}(t) \right] ,
\end{eqnarray}
where $\Theta(t,r)$ is dimensionless function,
\begin{eqnarray} \label{Theta}
\Theta(t,r)= \frac{m_f}{2\alpha \bar{t}} \int_{\bar{t}}^t
\frac{dt'}{\omega_{\bar{k}}(t')}+ \frac{\bar{k}r}{4}.
\end{eqnarray}
Note that $\Theta(t)=\Theta(t,r=0)$ becomes significant only after
$t-\bar{t}> \bar{t}$. With these we can write some general form of
the approximation of the integral,
\begin{eqnarray} \label{approx2}
\int dk g(k)k^2e^{2 \tilde{\theta}_{\bf k}(t_k)} e^{\pm
  i[2\theta^-_{\bf k}(t)\pm kr]}
\simeq \sqrt{\frac{\pi}{2}} \left(\frac{m_f}{\alpha
  \bar{t}}\right)^{3/2} e^{2\tilde{\theta}_{\bar{k}}(\bar{t})
    -2 \Theta^2(t,r)}
  g\left[\bar{k}\pm i\bar{k}\Theta(t,r)\right]
  e^{\pm i[2\theta^-_{\bar{k}}(t) \pm \bar{k}r]} .
\end{eqnarray}

\section{Rough estimation of $G$ integrals}
The integral $Q(m,t)$ in Eq.~(104)
\begin{eqnarray} \label{Gint}
\int_{|{\bf k}|>m_f} G^S_{\bf k}(t)- I_0(m^2)
 = \int_{|{\bf k}|>m_f} \frac{g^+_{\bf k}-1}{2
    \omega_{\bf k}(t)}
  -\int_{{\bf k}=0}^{m_f}\frac{1}{2\omega_{\bf k}}
-\int_{|{\bf k}|>m_f} \frac{\Omega_{-,\bf k}}{2\omega_{\bf k}}
   \cos 2 \theta^+_{\bf k}(t),
\end{eqnarray}
consists of three terms. The final term is unimportant to our
calculation and decreases as time so we ignore it. The second
integral is exactly integrable to give
\begin{eqnarray} \label{int2}
\int_{{\bf k}=0}^{m_f}\frac{1}{2\omega_{\bf
k}}=\frac{m_f^2}{8\pi^2}\left[\sqrt{1+\frac{x}{x_f}}-\frac{x}{x_f}
\ln \left(\sqrt{\frac{x_f}{x}}+ \sqrt{1+ \frac{x_f}{x}}\right)
\right].
\end{eqnarray}
The first integral of Eq.~(\ref{Gint}) cannot be exactly
integrable. Since the argument of the integral decreases rapidly,
see~(\ref{intermediate}), we series expand the integrand around
$k=m_f$:
\begin{eqnarray} \label{series:1}
\frac{g^+_{\bf k}-1}{2
    \omega_{\bf k}(t)}\sim
    \sqrt{\frac{m_i^2+m_f^2}{2m_f(m^2+m_f^2)}}
    \left[\frac{1}{2\sqrt{k-m_f}}-
    \sqrt{\frac{2m_f}{m_i^2+m_f^2}}+ O(\sqrt{k-m_f}/m_f)\right],
\end{eqnarray}
Since $g_{\bf k}^+-1$ is always positive definite, the above
series breakdown at the point, $\displaystyle
k'=m_f+\frac{m_i^2+m_f^2}{8m_f} $, where the series becomes zero.
Therefore, we estimate the first integral by
\begin{eqnarray} \label{int:1}
\int_{|{\bf k}|>m_f} \frac{g^+_{\bf k}-1}{2
    \omega_{\bf k}(t)}
 &\sim& \sqrt{\frac{m_i^2+m_f^2}{2m_f(m^2+m_f^2)}}
    \frac{1}{2\pi^2}\int_{m_f}^{k'} dk k^2
     \left[\frac{1}{2\sqrt{k-m_f}}-
    \sqrt{\frac{2m_f}{m_i^2+m_f^2}}\right] \\
&=&\frac{1}{240\pi^2} \frac{m_i^2+m_f^2}{m_f\sqrt{m^2+m_f^2}}
\left[{k'}^2+ 3k'm_f +11m_f^2\right]. \nonumber
\end{eqnarray}
In summary,
\begin{eqnarray} \label{GS-I}
\int_{|{\bf k}|>m_f} G^S_{\bf k}(t)- I_0(m^2)
 \sim \frac{m_f^2g(x_i,x_f)}{\sqrt{x/x_f+1}}-\frac{m_f^2}{8\pi^2}
  \left[\sqrt{1+\frac{x}{x_f}}-\frac{x}{x_f}
\ln \left(\sqrt{\frac{x_f}{x}}+ \sqrt{1+ \frac{x_f}{x}}\right)
\right],
\end{eqnarray}
where $\displaystyle g(x_i,x_f)=\frac{1}{240\pi^2}
\left(\frac{x_i}{x_f}+1\right)\left[\frac{{k'}^2}{m_f^2}+ 3
\frac{k'}{m_f}+ 11 \right]$.

Similarly, we may obtain
\begin{eqnarray} \label{g+-1-I}
&&\frac{1}{2}\int_{|{\bf k}|>m_f}\omega_{\bf k}(t)(g^+_{\bf
  k}-1)-\frac{(m_i^2+m_f^2)^2}{8}\bar{I}_{-1}(m^2,m_f)
  =\frac{1}{2}\int_{|{\bf k}|>m_f}\omega_{\bf k}(t)
   \left[g^+_{\bf
  k}-1-\frac{(m_i^2+m_f^2)^2}{8\omega_{\bf k}^4(t)}\right] \\
&&~~\sim \frac{m_f^4}{32\pi^2}\sqrt{x/x_f+1} g'(x_i,x_f) ,
\nonumber
\end{eqnarray}
where $g'(x_i,x_f)$ is a positive number dependent on $x_i$ and
$x_f$ only. Since we only need its large $m$ dependence, we do not
explicitly write $g'$.

Other integrals which we need in calculating $V_{\tau}$ in
Eq.~(109) are given by
\begin{eqnarray} \label{g-term}
\frac{1}{2}\int_{|{\bf k}|<m_f} g_{\bf k}^- \omega_{\bf k} &=&
\frac{1}{4 \pi^2}\int_0^{m_f} dk  \omega_{\bf k}(t)\Omega_{+, \bf
k} k^2 e^{2 \tilde{\theta}_{\bf k}(t_k)} \simeq \frac{\bar{l}}{2}
\tilde{\Omega}_{+, \bar{k}}\omega_{\bar{k}}(t) , \\
\frac{1}{2}\int_{|{\bf k}|<m_f} \omega_{\bf k}
&=&\frac{m_f^4}{32\pi^2}\left[\sqrt{1+\frac{x}{x_f}}
 \left(2+\frac{x}{x_f}\right)-\frac{x^2}{x_f^2}\ln
 \left(\sqrt{\frac{x_f}{x}}+\sqrt{1+\frac{x_f}{x}}\right) \right],
\end{eqnarray}

\end{appendix}%

\vspace{4cm}

\end{document}